\UseRawInputEncoding

\documentclass[11pt]{article}
\usepackage{}
\usepackage{amssymb}
\usepackage{amsfonts}
\usepackage{mathrsfs}
\usepackage{graphicx}
\usepackage{amsmath}
\usepackage{color}
\usepackage{amsthm}
\usepackage{epstopdf}
\usepackage{subfigure}
\usepackage{pifont,bm}
\usepackage{tikz}
\usepackage{enumerate}
\usepackage{mathrsfs}
\usepackage{cite}
\usepackage{hyperref}
\usepackage{booktabs}
\usepackage{diagbox}
\usepackage{makecell}

\usepackage{appendix}

\theoremstyle{definition}

\makeatletter
\def\@biblabel#1{[#1]}
\makeatother

\makeatletter \@addtoreset{equation}{section}

\begin{document}
%\begin{CJK*}{GBK}{song}

\begin{titlepage}
\title{\bf{Lax pairs informed neural networks solving integrable systems
%\title{\bf{Lax pairs informed neural networks for studying integrable systems
\footnote{Corresponding authors.\protect\\
\hspace*{3ex} E-mail addresses: pu$\_$juncai@qq.com (J.C. Pu), ychen@sei.ecnu.edu.cn (Y. Chen)}
}}
\author{Juncai Pu$^{a}$, Yong Chen$^{b,c,*}$\\
%%%%%%%%%%%%%%%%%%%%%%%%%%%%%%%%%%%%%%%%%%%%%%%%%%%%%%%%%%%%%%%%%%%%%%%%%%%%%%%%%%%%%%%%%
%%%%%              以下两行为作者单位
%%%%%%%%%%%%%%%%%%%%%%%%%%%%%%%%%%%%%%%%%%%%%%%%%%%%%%%%%%%%%%%%%%%%%%%%%%%%%%%%%%%%%%%%%
\small \emph{$^{a}$Institute of Applied Physics and Computational Mathematics, Beijing, 100094, China}\\
\small \emph{$^{b}$School of Mathematical Sciences, Key Laboratory of Mathematics and Engineering}\\
\small \emph{Applications(Ministry of Education) $\&$ Shanghai Key Laboratory of PMMP, East}\\
\small \emph{China Normal University, Shanghai, 200241, China}\\
\small \emph{$^{c}$College of Mathematics and Systems Science, Shandong University of Science and}\\
\small \emph{Technology, Qingdao, 266590, China} \\
\date{}}
\thispagestyle{empty}
\end{titlepage}
\maketitle

\vspace{-0.5cm}
\begin{center}
\rule{15cm}{1pt}\vspace{0.3cm}

\parbox{15cm}{\small
{\bf Abstract}\\
\hspace{0.5cm}
Lax pairs are one of the most important features of integrable system. In this work, we propose the Lax pairs informed neural networks (LPNNs) tailored for the integrable systems with Lax pairs by designing novel network architectures and loss functions, comprising LPNN-v1 and LPNN-v2. The most noteworthy advantage of LPNN-v1 is that it can transform the solving of nonlinear integrable systems into the solving of a linear Lax pairs spectral problems, and it not only efficiently solves data-driven localized wave solutions, but also obtains spectral parameter and corresponding spectral function in Lax pairs spectral problems of the integrable systems. On the basis of LPNN-v1, we additionally incorporate the compatibility condition/zero curvature equation of Lax pairs in LPNN-v2, its major advantage is the ability to solve and explore high-accuracy data-driven localized wave solutions and associated spectral problems for integrable systems with Lax pairs. The numerical experiments focus on studying abundant localized wave solutions for very important and representative integrable systems with Lax pairs spectral problems, including the soliton solution of the Korteweg-de Vries (KdV) euqation and modified KdV equation, rogue wave solution of the nonlinear Schr\"odinger equation, kink solution of the sine-Gordon equation, non-smooth peakon solution of the Camassa-Holm equation and pulse solution of the short pulse equation, as well as the line-soliton solution of Kadomtsev-Petviashvili equation and lump solution of high-dimensional KdV equation. The innovation of this work lies in the pioneering integration of Lax pairs informed of integrable systems into deep neural networks, thereby presenting a fresh methodology and pathway for investigating data-driven localized wave solutions and Lax pairs spectral problems.

}

\vspace{0.5cm}
\parbox{15cm}{\small{

\vspace{0.3cm} \emph{Key words: Lax pairs informed neural networks; Integrable systems; Localized wave solutions; Lax pairs}
}}
%\emph{PACS numbers:}  02.30.Ik, 05.45.Yv, 07.05.Mh.} }
\end{center}
\vspace{0.3cm} \rule{15cm}{1pt} \vspace{0.2cm}

\section{Introduction}
As a unique category of nonlinear systems, integrable systems exhibit rich mathematical structures and distinctive properties, including Lax pairs, multiple soliton solutions, and infinite conservation laws, setting them apart from general nonlinear systems \cite{Scott2004}. The exploration of integrable systems holds significant importance with wide-ranging applications in fields such as mathematics, physics, engineering, and various interdisciplinary fields, as a crucial research branch in nonlinear science, its advancements are also intricately tied to the progress in computer science. In recent years, amid the various revolutions in computer hardware and software technology, deep learning methods have emerged as a formidable approach for solving partial differential equations (PDEs). The idea is to optimize by constructing appropriate loss functions and use deep neural network (NN) to approximate unknown functions. Various neural network methods have been proposed, such as the physics-informed neural network (PINN) \cite{Raissi-JCP-2019} and the deep Galerkin method \cite{Sirignano-JCP-2018} with the loss established on the $L^2$-residual of the PDEs, and the deep Ritz method \cite{Ew-CMS-2018} with the loss based on the Ritz formulation, and the asymptotic-preserving neural network \cite{JinS-JSC-2023} with the loss satisfied the asymptotic preserving scheme, and so on. For other methods, we refer the reader to \cite{ChenJR-JML-2022,Dong-CMAME-2021,LuL-NMI-2021,WangSF-CMAME-2021,LiHY-CCP-2023,PuJC-PD-2023}.

For decades, benefiting from the rapid development of integrable system theory, a wealth of integrable models endowed with excellent properties has been accumulated, and numerous localized wave solutions as well as graceful program algorithms have been obtained, their provide a continuous stream of data samples and prior information for studying integrable systems using deep learning technology. Recent years, a series of significant research achievement have been achieved on the data-driven forward and inverse problems of integrable systems [or nearly integrable systems] by applying PINN and its improved algorithms. Chen's research group introduced deep learning methods to study integrable systems, and they have been dedicated to establishing an efficient framework for integrable deep learning methods, resulting in rich research outcomes \cite{LiJ-CTP-2020}. On the one hand, various data-driven localized wave solutions have been predicted from small amount of initial/boundary data, among which the most significant work is the first successful learning and extraction of rogue wave solutions \cite{PuJC-CPB-2021}. Additionally, they also obtained higher-order rational solutions, periodic wave solutions and rogue waves on the periodic background wave, vector rogue waves and interaction solutions between rogue wave and soliton, as well as new kink-bell type solution \cite{PuJC-ND-2021,PuJC-CSF-2022,LinSN-PD-2023}. Furthermore, various complex integrable models have been studied by employing PINN and its improved algorithms, including nonlocal integrable systems, coupled integrable systems, high-dimensional integrable systems, and variable coefficients integrable systems \cite{PuJC-CSF-2022,ZhouHJ-ND-2023}. On the other hand, various strategies have been utilized to make multiple improvements to deep NN algorithms, including adaptive activation functions, parameter regularization, time domain piecewise, and transfer learning \cite{PuJC-ND-2021,PuJC-CNS-2023,PuJC-PD-2023}. Especially, the properties of integrable systems are used to design new deep NN models suitable for integrable systems with such properties, such as the two-stage PINN based on conserved quantities \cite{LinSN-JCP-2022} and the PINN based on Miura transformations \cite{LinSN-PD-2023}. Moreover, there are many important works using deep learning algorithms to study integrable systems, we refer the reader to \cite{ZhouZJ-CMA-2023,ZhouXF-ND-2023,FangY-CSF-2022,YangXY-EPJP-2022,MoYF-PLA-2022}.

The inverse scattering method (IST) is one of the most important discoveries in modern mathematical physics \cite{Novikov-M-1984}, which was originally proposed by Gardner, Greene, Kruskal and Miura (GGKM) in 1967 to study the initial value problem of rapid decay of the Korteweg-de Vries (KdV) equation \cite{Gardner-PRL-1967}. In 1968, inspired by GGKM's work, Lax proposed the concept of Lax pairs and creatively proposed a general framework for using Lax pairs to represent integrable systems \cite{Lax-CPAM-1968}. In 1972, Zakharov and Shabat presented Lax pairs in matrix form for the nonlinear Schr\"{o}dinger (NLS) equation, and utilized IST to study the inverse scattering transformation and exact solutions of the NLS equation \cite{Zakharov-SPJ-1972}. Later, Ablowitz, Kaup, Newell and Segur proposed a class of integrable systems with a unified Lax pairs form, known as AKNS systems, and established a universal framework for inverse scattering theory for the initial-value problem of AKNS systems in 1974 \cite{Ablowitz-STA-1974,Ablowitz-PRL-1973}. Subsequently, many classic integrable systems with Lax pairs were proven to be solvable via IST \cite{Wadati-JPSJ-1973,Ablowitz-PRL2-1973,Ablowitz-SAM-1983}, which was also considered a milestone in the integrable systems theory and established the important position of integrable systems in many fields of mathematics and physics. Therefore, Lax pairs provides a theoretical basis for constructing a general framework of inverse scattering theory for integrable systems, promoting the development of integrable system theory.

Lax pairs are a key characteristic and important tool in the integrable systems theory, which are closely related to the integrability, conservation laws, classification, and exact solution solving of integrable systems. Lax pairs were first introduced by Lax in 1968, who stated that soliton equations can be represented by Lax pairs and Lax equations \cite{Lax-CPAM-1968}. Given a linear operator $L$ involved space $\bm{\mathrm{x}}$ and potential, let it satisfies spectral equation
\begin{align}\label{I1}
L\phi=\lambda\phi,
\end{align}
here $\phi$ is eigenfunction, $\lambda$ is spectral parameter. Consider the isospectral problem of $\lambda$ that is independent of time $t$, i.e. $\lambda_t=0$. Given an operator $A$, $\phi$ satisfies the linear equation
\begin{align}\label{I2}
\phi_t=A\phi.
\end{align}
If $\phi$ is required to satisfy both Eqs. \eqref{I1} and \eqref{I2}, then $L$ and $A$ satisfy the following operator equation
\begin{align}\label{I3}
L_t-[A,L]=0,
\end{align}
where $[A,L]=AL-LA$. Then Eq. \eqref{I3} is called the Lax equation, and the spectral problems \eqref{I1} and \eqref{I2} are called the Lax pairs. Once given a specific operator $L$, one can obtain the specific compatibility condition equation
\begin{align}\label{I3.5}
f_{\rm cce}: \phi_{\bm{\mathrm{x}}t}-\phi_{t\bm{\mathrm{x}}}=0/\phi_{\bm{\mathrm{x}}\bm{\mathrm{x}}t}-\phi_{t\bm{\mathrm{x}}\bm{\mathrm{x}}}=0.
\end{align}
Certainly, different integrable systems correspond to different compatibility condition equations, and here we only provide a unified representation Eq. \eqref{I3.5}. Generally, compatibility condition equation \eqref{I3.5} is equivalent to Lax equation \eqref{I3}, both of which can derive the desired soliton equation. In addition to the above operator representations, Lax pairs can also be expressed in matrix form, usually Lax pairs with operator form can be rewritten as Lax pairs with matrix form. Specially, as we consider the case where space $\bm{\mathrm{x}}=x$ is one-dimensional, then let a pair of spectral problems \eqref{I1} and \eqref{I2} be rewritten as
\begin{align}\label{I4}
\begin{split}
&\Phi_x=M\Phi,\\
&\Phi_t=N\Phi.
\end{split}
\end{align}
Where $\Phi$ is an $n$-dimensional column vector $\Phi=(\Phi_1,\Phi_2,\cdots,\Phi_n)^T$, $M$ and $N$ are $n$-th order matrices that depend on the potential $\bm{q}=\bm{q}(x,t)$ and spectral parameter $\lambda$. If requirements Eqs. \eqref{I4} are compatible, that is $\Phi$ satisfies compatibility condition equation $\Phi_{xt}-\Phi_{tx}=0$, then we can derive that $M$ and $N$ must be satisfied zero curvature equation
\begin{align}\label{I5}
f_{\rm zce}: M_t-N_x+[M,N]=0.
\end{align}
Usually, we refer to Eqs. \eqref{I4} as Lax pairs, thus the compatibility condition equation \eqref{I3.5} is equivalent to the zero curvature equation \eqref{I5}. The corresponding integrable systems can be derived from both the compatibility condition equation \eqref{I3.5} and zero curvature equation \eqref{I5}. If a PDE can be generated via the compatibility condition \eqref{I3} for operator Lax pairs \eqref{I1}-\eqref{I2} or zero curvature equation \eqref{I5} for matrix Lax pairs \eqref{I4}, then this PDE is said to be Lax integrable. Lax pairs are a prominent feature in the integrable systems theory, which can simplify the dynamic equations of integrable systems, thereby helping researchers study the integrability, construct conserved quantities, and classify integrable families. Furthermore, some methods for solving exact solutions of integrable systems also rely on the Lax pairs of integrable systems, such as the Darboux transformation method \cite{Matveev-SV-1991}. In summary, the proposal of Lax pairs greatly promoted the development of exactly solving methods of integrable systems and established its important position in integrable system theory, mathematical physics, and other related fields. Hence, the successful application of Lax pairs in aforementioned fields reminds us whether Lax pairs can be successfully applied to the deep NN for studying data-driven problems of integrable systems?

We introduce Lax pairs and their compatibility conditions/zero curvature equation into deep NNs, and propose Lax pairs informed neural networks (LPNNs), the novel algorithm exhibit remarkable efficiency or high-accuracy in solving integrable systems. In order to fully demonstrate the high efficiency and high precision of LPNNs in solving integrable systems with Lax pairs, we utilized LPNNs to solve several important integrable systems and their Lax pairs spectrum problems. The KdV equation is the first PDE to describe solitary wave and the first integrable model studied by applying the IST \cite{Korteweg-PM-1895,Gardner-PRL-1967}, and the KdV equation is closely related to the origin of soliton and the flourishing development of integrable system theory, which has had a significant impact on the fields of nonlinear science and mathematical physics \cite{Zabusky-PRL-1965}, it is the most iconic model in the soliton and integrable systems theory. The Camassa-Holm (CH) equation is the earliest integrable model discovered to both possess non-smooth solutions and Lax pairs, which can describe the unidirectional water wave motion in shallow water waves and has important physical significance \cite{Camassa-PRL-1993}. The Kadomtsev-Petviashvili (KP) equation is a high-dimensional extension of the KdV equation, which describes the motion of two-dimensional water waves and has widely applied in the fields of fluid mechanics and theoretical physics \cite{Kadomtsev-DAN-1970}. The term ``breather'' arose from studies of the sine-Gordon (SG) equation, which is one of the most fundamental equations of integrable systems and has important applications in fields such as nonlinear optics, crystal dislocation, and superconductivity \cite{Frenkel-JP-1939}. The modified KdV (mKdV) equation is a KdV equation with cubic nonlinear term, and its transformation relationship with the solution of the KdV equation has opened up a research boom in Miura transformations \cite{Zabusky-NPDE-1967}. The NLS equation has important applications in various fields such as quantum mechanics, optics, plasmas, and Bose-Einstein condensates, and it is also the most fundamental equation for describing rogue wave phenomena \cite{Zakharov-SPJ-1972}. Unlike the KdV, mKdV, SG and NLS equations in the AKNS hierarchy, the short pulse (SP) equation is the most classical integrable model in Wadati-Konno-Ichikawa (WKI) equations \cite{Wadati-JPSJ-1979}, and possess WKI-type Lax pairs, which has important applications in many physical fields such as nonlinear optics \cite{Schafer-PD-2004}. These integrable models are the most significant, foundational and classic models in the realm of integrable systems, and many integrable models can be derived from the deformation and generalization of these basic integrable models. Hence, the successful application of LPNNs in solving these significant, classic and representative integrable systems serves as a compelling demonstration of the efficacy of the LPNNs algorithm.

The main highlights of this article are as follows:\\
\textbf{1.} We cleverly transform relatively complex linear integrable systems into simple linear equations [Lax pairs spectral problems] to improve the training efficiency of the network. LPNN-v1 relies exclusively on the Lax pairs of integrable systems to extract information, significantly simplifying the solving of complex integrable systems. LPNN-v1 can not only effectively solve data-driven localized wave solutions in such systems, but also conveniently learn the spectral parameter and their corresponding spectral function in Lax pairs spectral problems. The numerical results indicate that LPNN-v1 is applicable to both low-dimensional and high-dimensional integrable systems, and exhibits efficient training performance in solving smooth and non-smooth solutions. Specifically, for localized wave solution of certain integrable systems, LPNN-v1 can reduce training time by more than 5 times than standard PINN. Moreover, owing to the fact that the Lax pairs representation of high-dimensional integrable systems is frequently considerably simpler than the equations themselves, LPNN-v1 enjoys substantial advantages in solving high-dimensional integrable systems.\\
\textbf{2.} In the case of certain simple integrable systems or intricate localized wave solutions, LPNN-v1 tends to lose its efficiency. Therefore, building upon LPNN-v1, we further introduce the compatibility condition/zero curvature equation of Lax pairs to propose the LPNN-v2. The numerical results indicate that LPNN-v2 can solve high-precision data-driven localized wave solutions and spectral problems for all integrable systems with Lax pairs, and can even improve training accuracy by an order of magnitude when solving localized wave for certain integrable systems.

The structure of the paper unfolds as follows: In section 2, we present the innovative models of LPNNs. Section 3 provides a comprehensive display of numerical experiments conducted to validate the effectiveness of our proposed methods. We encapsulate our work and draw meaningful conclusions in section 4.

\section{Methodology}
Generally, we consider a multi-dimensional spatiotemporal real nonlinear integrable system with operator Lax pairs spectral problem \eqref{I1}-\eqref{I2} or matrix Lax pairs spectral problem \eqref{I4} in the general form given by
\begin{subequations}\label{M1}
\begin{equation}\label{M1a}
\mathcal{F}[\bm{q},\bm{q}^2,\cdots,\nabla_{t}\bm{q},\nabla^2_{t}\bm{q},\cdots,\nabla_{\bm{\mathrm{x}}}\bm{q},\nabla^2_{\bm{\mathrm{x}}}\bm{q},\cdots,\bm{q}\cdot\nabla_{t}\bm{q},\cdots,\bm{q}\cdot\nabla_{\bm{\mathrm{x}}}\bm{q},\cdots]=0,
\end{equation}
\begin{equation}\label{M1b}
f_{\rm{Lp}}:\,\left\{\begin{aligned}
L\phi=\lambda\phi\\
\phi_t=A\phi
\end{aligned}\right., \,\,\text{usually for $\bm{\mathrm{x}}\in\Omega$},\,\text{or}\,
\left\{\begin{aligned}
\Phi_{x}=M\Phi\\
\Phi_t=N\Phi
\end{aligned}\right.,\,\text{usually for $\bm{\mathrm{x}}=x$},
\end{equation}
\end{subequations}
in which potential $\bm{q}=\bm{q}(\bm{\mathrm{x}},t)\in\mathbb{R}^{n\times 1}$ is the $n$-dimensional latent solution, $\bm{\mathrm{x}}\in\Omega$ specifies the $n$-dimensional space and $t\in[T_\mathrm{i},T_\mathrm{f}]$ denotes time [$T_\mathrm{i}$ and $T_\mathrm{f}$ respectively indicate the initial time and final time], $\nabla$ is the gradient operator with respect to $\bm{\mathrm{x}}$ and $t$, $\mathcal{F}[\cdot]$ is a complex nonlinear operator of $\bm{q}$ and its spatiotemporal derivatives. Here linear operator $L$ involves space $\bm{\mathrm{x}}$ and potential $\bm{q}$, $\lambda$ indicates spectral parameter and $\phi$ represents spectral function corresponding to spectral parameter. $\Phi=\Phi(\bm{\mathrm{x}},t)$ stands for vector spectral function corresponding to spectral parameter $\lambda$ in matrixs $M$ and $N$. Usually, from Lax pairs \eqref{M1b}, we can derive all integrable systems by means of compatibility condition equation \eqref{I3.5} [for Lax pairs of operator form] and zero curvature equation \eqref{I5} [for Lax pairs of matrix form].

Then we consider the initial and boundary conditions of spatiotemporal nonlinear integrable system denoted by
\begin{align}\label{M2}
\begin{split}
&\qquad\mathcal{I}[\bm{q},\phi/\Phi;\bm{\mathrm{x}}\in\Omega,t=T_{\mathrm{i}}]=0,\\
&\mathcal{B}[\bm{q},\phi/\Phi,\nabla_{\bm{\mathrm{x}}}\bm{q};\bm{\mathrm{x}}\in\partial\Omega,t\in[T_\mathrm{i},T_\mathrm{f}]]=0.
\end{split}
\end{align}
If we consider a complex valued potential $\bm{\hat{q}}\in\mathbb{C}^{n\times 1}$ for nonlinear integrable systems, we can utilize decomposition $\bm{\hat{q}}=\bm{\hat{u}}+\mathrm{i}\bm{\hat{v}}$ to derive two real-value functions $\bm{\hat{u}}\in\mathbb{R}^{n\times 1}$ and $\bm{\hat{v}}\in\mathbb{R}^{n\times 1}$, then back to the problem of Eq. \eqref{M1a}. The initial and boundary points set $\mathcal{D}_{\mathrm{ib}}$ for training are sampled randomly through corresponding initial and boundary conditions \eqref{M2}, and the collocation points set $\mathcal{D}_{\mathrm{c}}$ for training are generated by the Latin Hypercube Sampling method \cite{Stein-T-1987}.

The next part, we utilize the Lax pairs, a prominent feature of integrable systems, to design novel deep learning method, namely LPNNs. The NN part of LPNNs still adopts fully connected networks, while the Lax pairs informed part is constructed through Lax pairs, compatibility condition equation/zero curvature equation. Given an input $\bm{\mathrm{x}}'$ [for convenience, here $\bm{\mathrm{x}}'$ include $\bm{\mathrm{x}}$ and $t$], the output of the $L$-layer deep feedforward NN $F$ is a combination of the affine transformations $\{\mathcal{A}_d\}^L_{d=1}$ and nonlinear activation functions $\{\sigma_d\}^{L-1}_{d=1}$, stated as follows
\begin{align}\nonumber
\begin{split}
&F(\bm{\mathrm{x}}';\bm{\theta})=\mathcal{A}_L\circ\sigma_{L-1}\circ\mathcal{A}_{L-1}\cdots\circ\sigma_2\circ\mathcal{A}_2\circ\sigma_1\circ\mathcal{A}_1(\bm{\mathrm{x}}'),\\
&\mathcal{A}_d(\bm{\mathrm{x}}'_{d-1})\triangleq\bm{W}^d\bm{\mathrm{x}}'_{d-1}+\bm{b}^d,\,\,\bm{\mathrm{x}}'_{0}:=\bm{\mathrm{x}}',\\
&\bm{\mathrm{x}}'_{d}=\sigma_d\circ\mathcal{A}_d\triangleq\sigma_d\big(\mathcal{A}_d(\bm{\mathrm{x}}'_{d-1})\big),
\end{split}
\end{align}
in which the weights and bias term are $\bm{W}^{d}\in\mathbb{R}^{N_d\times N_{d-1}}$ and $\bm{b}^d\in\mathbb{R}^{N_d}$ in the NN associated with the $d$-th layer, and the $N_d$ denote the number of neurons contained in $d$-th hidden layer. here popular choices of $\sigma_d$ include the ReLU function, the sigmoid function, the hyperbolic tangent function, and so on. In this work, we all adopt the hyperbolic tangent as activation function. the set of trainable parameters $\bm{\theta}\in\mathcal{N}$ consists of $\big\{\bm{W}^d,\bm{b}^d\big\}_{d=1}^{D}$, in which $\mathcal{N}$ is the parameter space.

\begin{figure}[htbp]
\centering
\begin{minipage}[t]{0.99\textwidth}
\centering
\includegraphics[height=7cm,width=14cm]{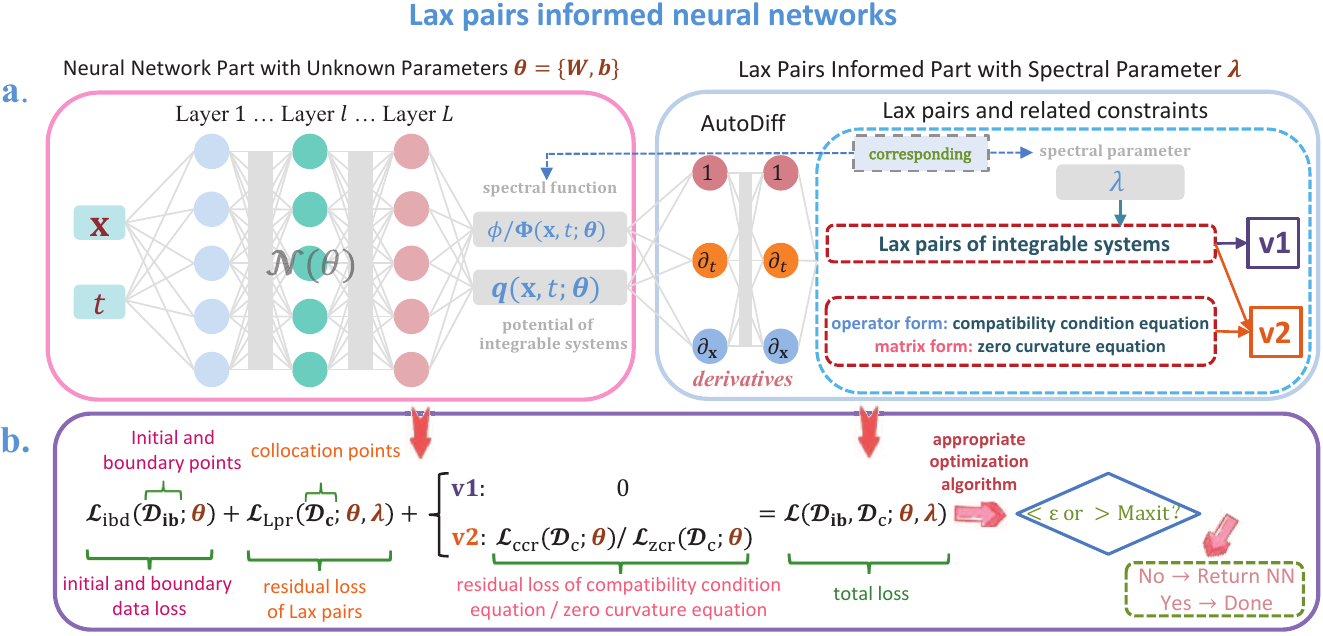}
%\caption{fig1}
\end{minipage}
\centering
\caption{(Color online) Schematic architecture of the LPNNs model for integrable systems with Lax pairs, including LPNN-v1 and LPNN-v2. $\bm{\mathrm{a}}$ the NN part (left part) and Lax pairs informed part (right part). $\bm{\mathrm{b}}$ the components of the loss function and optimize process in LPNNs.}
\label{LPNN-v12}
\end{figure}

For the construction of Lax pairs informed part of LPNN, we mainly divide it into two methods: LPNN-v1 only uses Lax pairs as Lax pairs informed; while LPNN-v2 utilizes Lax pairs and their zero curvature equation/compatibility condition equation as Lax pairs informed. Each of the two schemes has its unique advantages, among which the LPNN-v1 relies entirely on the Lax pairs of integrable systems to extract Lax pairs informed, greatly simplifying the solving of complex integrable systems, it is well-suited for studying localized wave solutions and spectral problems of complex integrable systems with uncomplicated Lax pairs, and it stands out for its high training efficiency. While the LPNN-v2 method is primarily tailored for delving into various localized wave solutions and spectral problems within all integrable systems with Lax pairs, and it usually attains high training accuracy. Fig. \ref{LPNN-v12} depicts the schematic architecture of the LPNNs model, encompassing both the LPNN-v1 and LPNN-v2.

Specifically, Fig. \ref{LPNN-v12} displays two versions of LPNNs through the upper and lower panels, namely LPNN-v1 and LPNN-v2. The upper panel Fig. \ref{LPNN-v12} $\bm{\mathrm{a}}$ showcase the NN part and Lax pairs informed part, where the left part of panel Fig. \ref{LPNN-v12} $\bm{\mathrm{a}}$ is the fully connected deep feedforward network, the input $\bm{\mathrm{x}}'$ include $\bm{\mathrm{x}}$ and $t$, the output $F$ contains potential $\bm{q}$ and spectral function $\phi/\Phi$. While the right part of panel Fig. \ref{LPNN-v12} $\bm{\mathrm{a}}$ is Lax pairs informed network dominated by the Lax pairs and their related conditions. Different Lax pairs constraints correspond to different versions of LPNNs, that is LPNN-v1 only corresponds to ``Lax pairs of integrable systems'', then LPNN-v2 corresponds to ``Lax pairs of integrable systems'' and corresponding ``compatibility condition equation''/``zero curvature equation''. The spectral parameter $\lambda$ correspond to the spectral function $\Phi$, which can be set as hyper-parameter or trained through network, and is reflected in the Lax pairs of integrable systems. The Lax pairs of certain high-dimensional integrable systems may not contain spectral parameter, we do not need to consider the spectral parameter term in this case. The left and right parts of panel Fig. \ref{LPNN-v12} $\bm{\mathrm{a}}$ are connected using automatic differentiation [AutoDiff]. The loss function and optimize process of LPNNs are depicted in the lower panel Fig. \ref{LPNN-v12} $\bm{\mathrm{b}}$, where the training parameters of total loss $\mathcal{L}$ are provided from two-part of panel Fig. \ref{LPNN-v12} $\bm{\mathrm{a}}$, the optimization algorithms appropriately selected Adam and/or L-BFGS optimizer for training of LPNNs. Ultimately, employ appropriate optimization algorithms to seek optimal parameters $\bm{\theta}^*$, this endeavor aims to minimize the total loss function $\mathcal{L}$ effectively. the total loss function is defined as

\begin{subequations}\label{M3}
\begin{equation}\label{M3a}
\text{LPNN-v1}: \mathcal{L}(\mathcal{D}_{\rm{ib}},\mathcal{D}_{\rm{c}};\bm{\theta},\lambda)=\mathcal{L}_{\rm{ibd}}(\mathcal{D}_{\rm{ib}};\bm{\theta})+\mathcal{L}_{\rm{Lpr}}(\mathcal{D}_{\rm{c}};\bm{\theta},\lambda),
\end{equation}
\begin{equation}\label{M3b}
\text{LPNN-v2}: \mathcal{L}(\mathcal{D}_{\rm{ib}},\mathcal{D}_{\rm{c}};\bm{\theta},\lambda)=\mathcal{L}_{\rm{ibd}}(\mathcal{D}_{\rm{ib}};\bm{\theta})+\mathcal{L}_{\rm{Lpr}}(\mathcal{D}_{\rm{c}};\bm{\theta},\lambda)+\left\{\begin{aligned}
&\text{operator form}:\,\mathcal{L}_{\rm{ccr}}(\mathcal{D}_{\rm{c}};\bm{\theta})\\
&\text{matrix form}:\,\mathcal{L}_{\rm{zcr}}(\mathcal{D}_{\rm{c}};\bm{\theta})
\end{aligned}\right.,
\end{equation}
\end{subequations}
Here, $\mathcal{L}_{\rm{ibd}}$ represents the initial and boundary data loss, $\mathcal{L}_{\rm{Lpr}}$ indicates residual loss of Lax pairs, then $\mathcal{L}_{\rm{ccr}}$ and $\mathcal{L}_{\rm{zcr}}$ respectively represent the residual loss of compatibility condition equation [for Lax pairs of operator form] and residual loss of zero curvature equation [for Lax pairs of matrix form]. they can be defined as following
\begin{align}\label{M4}
&\mathcal{L}_{\rm{ibd}}(\mathcal{D}_{\rm{ib}};\bm{\theta})=\frac{1}{N_{\rm{ib}}}\big\|\bm{q}^{\bm{\theta},\mathrm{ib}}-\bm{q}^{\bm{m},\mathrm{ib}}\big\|^2_2,\\
\label{M5}
&\mathcal{L}_{\rm{Lpr}}(\mathcal{D}_{\rm{c}};\bm{\theta},\lambda)=\frac{1}{N_{\rm{c}}}\big\|f^{\rm{c}}_{\rm{Lp}}\big\|^2_2,
\end{align}
and
\begin{align}\label{M6}
&\mathcal{L}_{\rm{ccr}}(\mathcal{D}_{\rm{c}};\bm{\theta})=\frac{1}{N_{\mathrm{c}}}\big\|f^{\rm{c}}_{\rm cce}\big\|^2_2,\\
\label{M7}
&\mathcal{L}_{\rm{zcr}}(\mathcal{D}_{\rm{c}};\bm{\theta})=\frac{1}{N_{\mathrm{c}}}\big\|f^{\rm{c}}_{\rm zce}\big\|^2_2,
\end{align}
where $N_{\rm{ib}}$ and $N_{\rm{c}}$ represent respectively the number of elements in sets $\mathcal{D}_{\mathrm{ib}}$ and $\mathcal{D}_{\mathrm{c}}$, $\|\cdot\|_2$ denotes the $L^2$ norm. Then $\bm{q}^{\bm{\theta},\mathrm{ib}}$ represents the learning results of $\bm{q}^{\bm{\theta}}$ acting on initial and boundary points set $\mathcal{D}_{\mathrm{ib}}$. Besides, $\bm{q}^{\bm{m},\mathrm{ib}}$ represents the measurement data of $\bm{q}$ on initial and boundary points set $\mathcal{D}_{\mathrm{ib}}$. The $f^c_{\rm{Lpr}}$ is value of Lax pairs $f_{\rm{Lpr}}$ on collocation points set $\mathcal{D}_{\mathrm{c}}$. The $f^{\rm{c}}_{\rm cce}$ is value of compatibility condition equation $f_{\rm cce}$ on collocation points set $\mathcal{D}_{\mathrm{c}}$, and the $f^{\rm{c}}_{\rm zce}$ is value of zero curvature equation $f_{\rm zce}$ on collocation points set $\mathcal{D}_{\mathrm{c}}$. Finally, we summarize the main steps of LPNNs in Algorithm \ref{Tab:LPNNs}.

\begin{table}[htbp]
  \label{Tab:LPNNs}
  \centering
  \begin{tabular}{p{15cm}}
  \toprule[2pt]
  Algorithm \ref{Tab:LPNNs}: The Lax pairs informed neural networks for integrable systems.  \\
  \midrule[2pt]
  \quad \textbf{Step 1}: Specification of training set in computational domain:\\
  \quad \emph{initial and boundary training points}: $\mathcal{D}_{\mathrm{ib}}$, \emph{collocation training points}: $\mathcal{D}_{\mathrm{c}}.$\\
  \quad \textbf{Step 2}: Construct neural network $F$ [including potential $\bm{q}$ and spectral function $\bm{\Phi}$] with random initialization of parameters $\bm{\theta}$ [contain/preset spectral parameter $\lambda$].\\
  \quad \textbf{Step 3}: Construct the Lax pairs informed part by substituting surrogate $F(\bm{\mathrm{x}}';\bm{\theta})$ into the Lax pairs of integrable systems in LPNN-v1 [the Lax pairs of integrable systems and compatibility condition equation/ zero curvature equation in LPNN-v2].\\
  \quad \textbf{Step 4}: Specification of the total loss function $\mathcal{L}(\mathcal{D}_{\rm{ib}},\mathcal{D}_{\rm{c}};\bm{\theta},\lambda)$.\\
  \quad \textbf{Step 5}: Seek the optimal parameters $\bm{\theta}^*$ using a suitable optimization algorithms for minimizing the total loss function $\mathcal{L}$ as\\
  \qquad\qquad\qquad\qquad\qquad\qquad\qquad\quad $\bm{\theta}^*=\mathop{\mathrm{arg\,min}}\limits_{\bm{\theta}\in\mathcal{N}}\mathcal{L}(\bm{\theta})$.\\
  \bottomrule[2pt]
  \end{tabular}
\end{table}

\section{Numerical Experiment}
In this section, we utilize three different types of deep learning models to study data-driven localized wave solutions, and solve Lax pairs spectral problems for integrable systems with several different types of Lax pairs, then provide detailed numerical results and related dynamic behavior figures, and compare them with other deep learning methods. Uniformly, in this work, all deep learning methods [involve the PINN and LPNNs] both possess 5 hidden-layer NNs with 100 neurons per hidden layer, namely $L=6$ and $N_d=100$. Additionally, the numerical results in this paper both are derived based on the TensorFlow-cpu 1.15 version, thus reader if utilize the TensorFlow-gpu 2.X version for training would result in significantly shorter training times in each cases.

\subsection{Efficient training performance of LPNN-v1}
In this subsection, we apply LPNN-v1 to study integrable systems with simple Lax pairs, and efficiently solve data-driven solutions and corresponding Lax pairs spectral problems.

$\bullet$ \textbf{Case 1: Korteweg-de Vries equation}

The KdV equation is a PDE that describes the motion of water waves, and is one of the most classic equations in the soliton theory and integrable systems. In 1834, British scientist Russell occasionally observed a type of water wave [also known as a solitary wave] whose shape and velocity did not change during the process of traveling, but he did not find an appropriate model or specific mathematical expression to describe this interesting natural phenomenon at the time. Until 1895, Korteweg and de Vries proposed the KdV equation and pointed out that the traveling wave solution of the equation could perfectly explain the solitary wave phenomenon discovered by Russell \cite{Korteweg-PM-1895}. In 1965, Zabusky and Kruskal first discovered the connection between the FPU problem proposed by Fermi, Pasta, Ulam, and Tsingou in 1955 and the KdV equation by utilizing numerical calculation methods \cite{Fermi-LANM-1955}, and explained the FPU regression phenomenon using the dynamic behavior of solitary waves in the KdV equation \cite{Zabusky-PRL-1965}. Since then, Zabusky and Kruskal proposed the term ``soliton'' in 1965 to reflect the properties of solitary waves and particle interactions \cite{Zabusky-PRL-1965}. Solitons are nonlinear localized waves that can maintain their shape and velocity during propagation, even when interacting with other waves. Therefore, the KdV equation is closely related to the origin of solitons and the flourishing development of integrable system theory, which has had a significant impact on the fields of nonlinear science and mathematical physics. Until now, the KdV equation is widely recognized as a paradigm for the description of weakly nonlinear long waves in many branches of physics and engineering.

We consider the KdV equation with Lax pairs [operator form] as follows
\begin{align}\label{Ne-v1-KdV-1}
&u_t+6uu_x+u_{xxx}=0,\\
\label{Ne-v1-KdV-2}
&f_{\rm{Lp}}:\,\bigg\{\begin{aligned}
&\phi_{xx}=(\lambda-u)\phi\\
&\phi_t=u_x\phi-(4\lambda+2u)\phi_{xxx}
\end{aligned},
\end{align}
thus KdV equation \eqref{Ne-v1-KdV-1} can be derived by means of Lax pairs \eqref{Ne-v1-KdV-2} and specific formula $\phi_{xxt}-\phi_{txx}=0$ of compatibility condition equation \eqref{I3.5}, where $\phi_{xxt}$ is computed from the first equation of Eq. \eqref{Ne-v1-KdV-2} and $\phi_{txx}$ is given by the second equation of Eq. \eqref{Ne-v1-KdV-2}. Then we consider following initial condition $\mathcal{I}$ and boundary conditions $\mathcal{B}$ for KdV equation \eqref{Ne-v1-KdV-1} in spatiotemporal region $[-5,5]\times[-5,5]$
\begin{align}\label{Ne-v1-KdV-3}
\begin{split}
&u(x,t=-5)=2\mathrm{sech}(20+x)^2,\,x\in[-5,5],\\
&u(-5,t)=2\mathrm{sech}(-4t-5)^2,\,u(5,t)=2\mathrm{sech}(-4t+5)^2,\,t\in[-5,5].
\end{split}
\end{align}
For the Lax pairs of KdV equation \eqref{Ne-v1-KdV-1}, spectral function $\phi(x,t)\in\mathbb{R}^{1\times 1}$ satisfied free initial-boundary condition and initialized to $\phi(x,t)=0$ in spatiotemporal region $[-5,5]\times[-5,5]$. Then we apply conventional PINN and novel LPNN-v1 to solve the KdV equation \eqref{Ne-v1-KdV-1} with spectral problem \eqref{Ne-v1-KdV-2}, and set spectral parameter $\lambda=1$ as well as training points $N_{\mathrm{ib}}=400,\,N_{\mathrm{c}}=10000$. the relative $L^2$ norm error of the LPNN-v1 model achieves 5.756885$\rm e$-03 for data-driven single-soliton solution $u(x,t)$ in 64.6284 seconds, and the number of iterations is 327.

\begin{figure*}[!htbp]
\centering
\subfigure[]{\includegraphics[height=1.0in,width=1.6in]{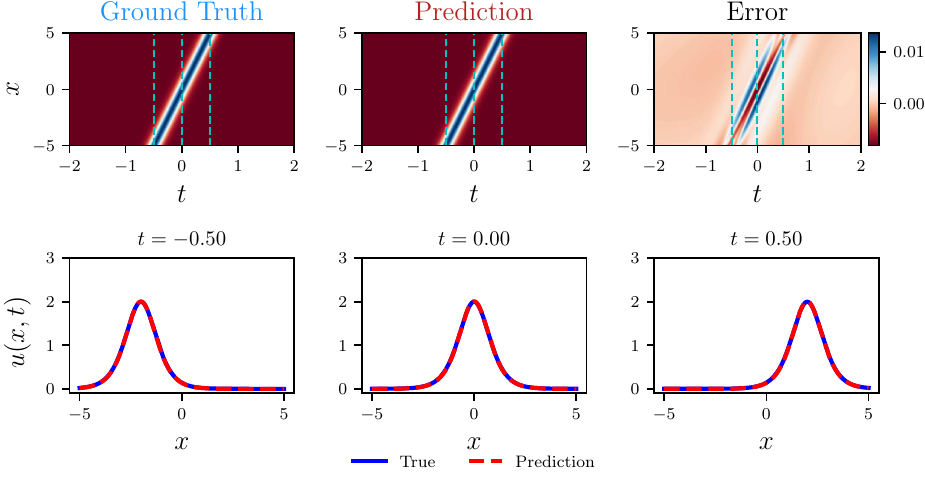}}\hspace{0.15cm}
\subfigure[]{\includegraphics[height=1.0in,width=1.4in]{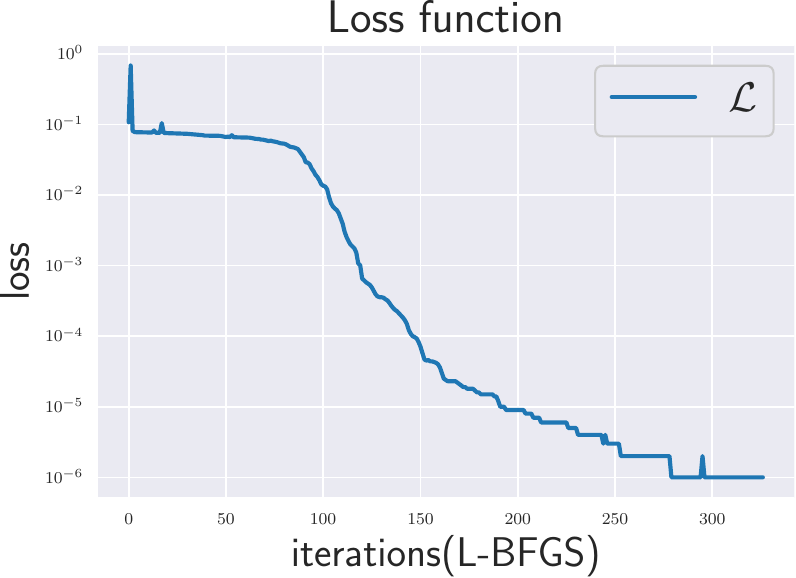}}\hspace{0.15cm}
\subfigure[]{\includegraphics[height=1.0in,width=1.2in]{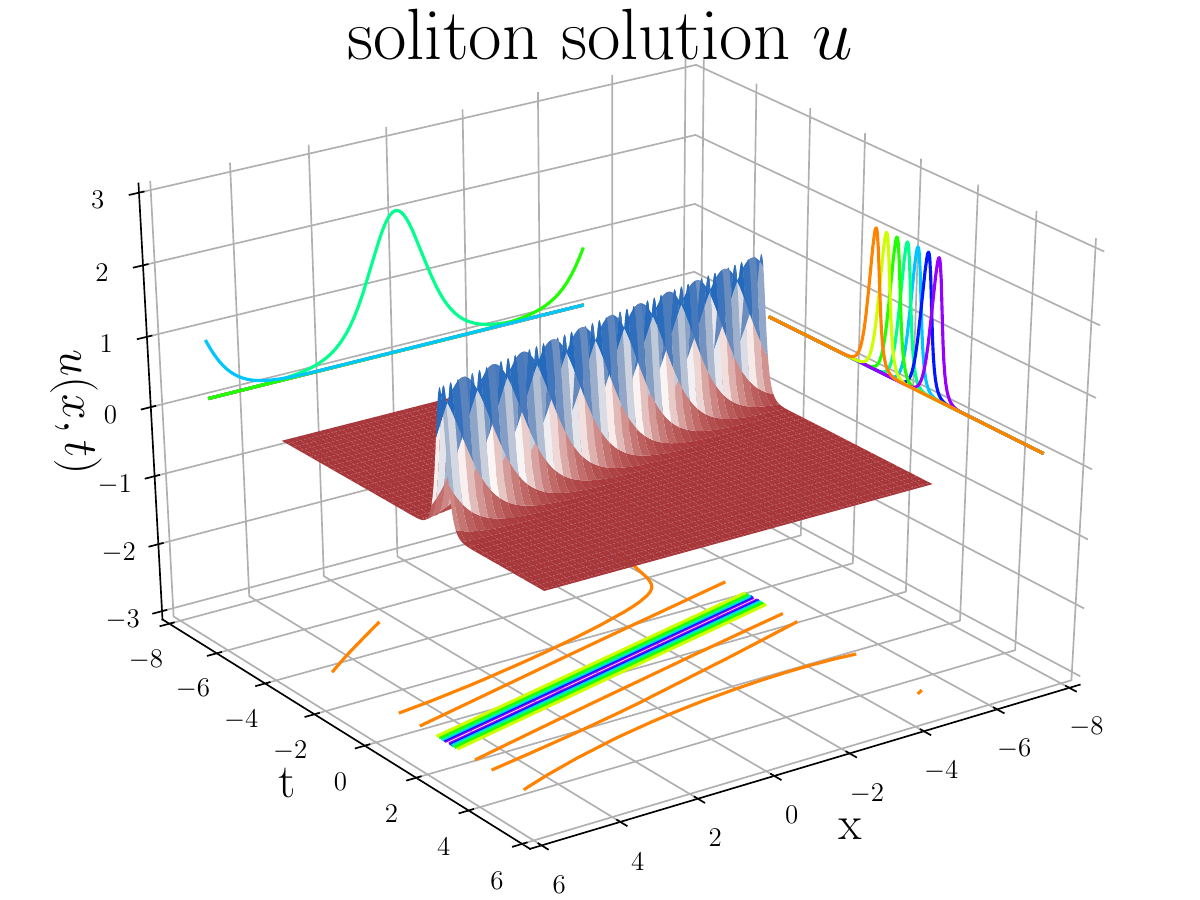}}
\subfigure[]{\includegraphics[height=1.0in,width=1.2in]{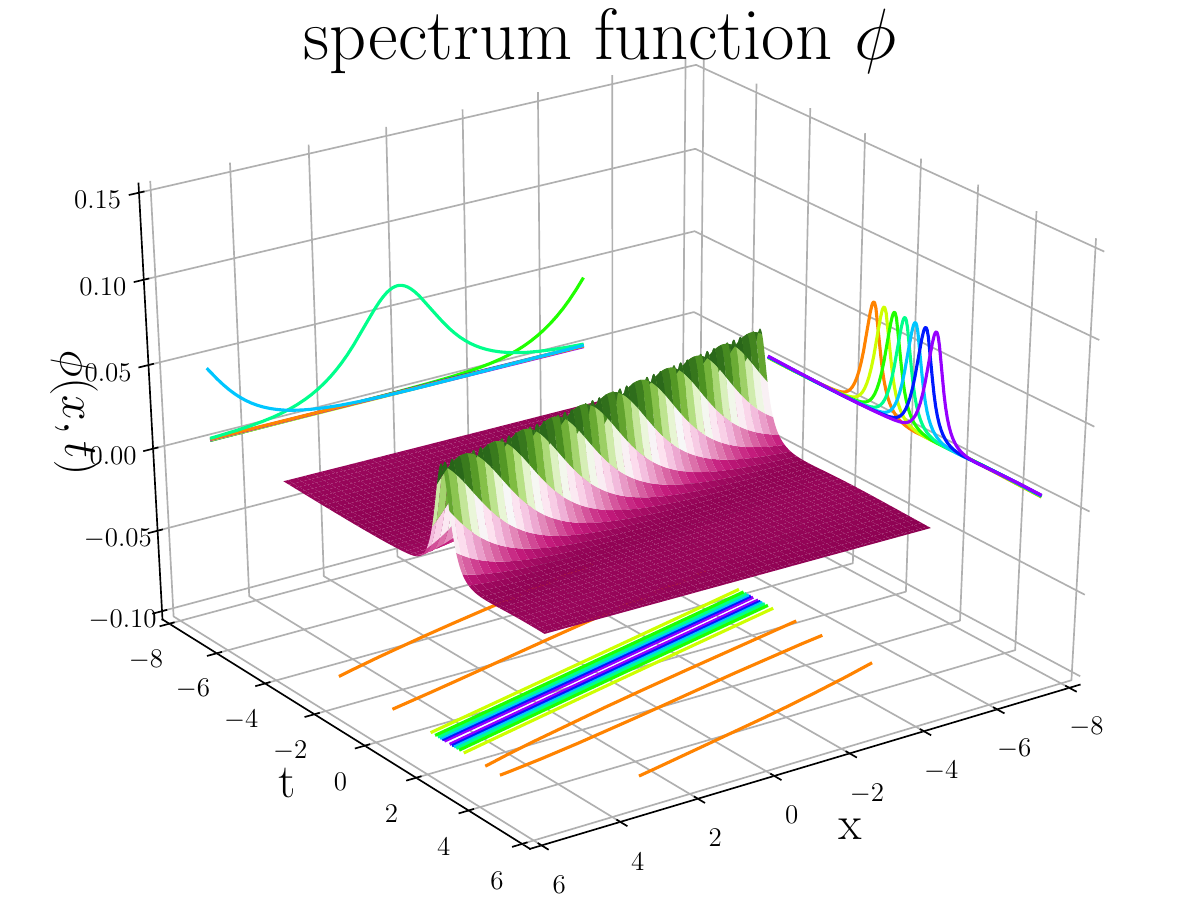}}
\caption{(Color online) The training results of single-soliton solution $u(x,t)$ and spectral function $\phi(x,t)$ for KdV equation arising from the LPNN-v1. (a) The ground truth, prediction and error dynamics density plots, as well as sectional drawings which contain the true and prediction single-soliton solution at three distinct moments $t=-0.5, 0, 0.5$; (b) Evolution graph of the loss function in LPNN-v1; (c) The three-dimensional plot with contour map for the data-driven single-soliton solution; (d) The three-dimensional plot with contour map for the data-driven spectral function corresponding to spectral parameter $\lambda=1$. \label{figLPNN-v1-KdV}}
\end{figure*}

Fig. \ref{figLPNN-v1-KdV} manifests the deep learning results of the data-driven single-soliton solution $u(x,t)$ and spectral function $\phi(x,t)$ stemming from the LPNN-v1 for
the KdV equation \eqref{Ne-v1-KdV-1}. In Fig. \ref{figLPNN-v1-KdV}(a), we display the density plots of the true dynamics, prediction dynamics and error dynamics, then showcase its corresponding amplitude scale size on the right side of density plots, and exhibit the sectional drawings which contain the learned and true solution at three different moments. The evolution curve figures of the loss function arising from the LPNN-v1 are displayed in Fig. \ref{figLPNN-v1-KdV}(b). Figs. \ref{figLPNN-v1-KdV}(c) and (d) indicate respectively the three-dimensional plot with contour map on three planes for the predicted single-soliton solution $u(x,t)$ and learned spectral function $\phi(x,t)$.

To further showcase the efficiency of LPNN-v1 in studying localized wave solutions of integrable systems with concise Lax pairs, we provide a comparison of the training errors and training time between LPNN-v1 and conventional PINNs in Table \ref{TabLPNN-v1-KdV}, which spans same spatio-temporal domain as well as possesses identical hyperparameters and network setting. From Table \ref{TabLPNN-v1-KdV}, we can obtain that although the training error using LPNN-v1 is slightly greater than that using standard PINN, the training time of LPNN-v1 is much shorter than that required using PINN, which benifit from the simpler Lax pairs form of the KdV equation compared to the KdV equation itself. In realistic problems, faster and efficient network training is often more favored when the training errors of the network are not significantly different. Therefore, the numerical results in this subsection indicate that LPNN-v1 is particularly suitable for studying cases where the integrable system model itself is complicated but possesses simple Lax pairs when training certain data-driven localized wave solutions.

\begin{table}[htbp]
  \caption{Performance comparison between LPNN-v1 and conventional PINN for solving KdV equation}
  \label{TabLPNN-v1-KdV}
  \centering
  \scalebox{0.8}{
  \begin{tabular}{l|c|c|c|c|c|c}
  \toprule
  \small{\textbf{Types}} & $\bm{\mathrm{x}}\times t$ & $\mathcal{D}_{\mathrm{ib}},\mathcal{D}_{\mathrm{c}}$ & optimizer & $\lambda$ & $L^2$ norm error & training time \\
  \hline
  PINN   & [-5,5]$\times$[-5,5] & 400,10000 & L-BFGS & $\backslash$ & 2.742974$\rm e$-03 & 200.5237s \\
  \hline
  LPNN-v1 & [-5,5]$\times$[-5,5] & 400,10000 & L-BFGS & 1.0 & 5.756885$\rm e$-03 & 64.6287s \\
  \bottomrule
  \end{tabular}}
\end{table}

$\bullet$ \textbf{Case 2: Camassa-Holm equation}

For verifying the effectiveness of LPNN-v1 in solving non-smooth solutions of integrable systems, we considered the classical CH equation
\begin{align}\label{Ne-v1-CH-1}
u_t-u_{xxt}+2\omega u_x+3uu_x-2u_xu_{xx}-uu_{xxx}=0,
\end{align}
arises as a model describing the unidirectional propagation of shallow water waves over a flat bottom, where $u(x,t)$ represent the height of the free surface about a flat bottom and the real constant $\omega$ being related to the critical shallow water speed. The CH equation was first appeared in Ref. \cite{Fokas-PD-1981} as an abstract bi-Hamiltonian equation with infinitely many conservation laws. Camassa and Holm subsequently discovered Eq. \eqref{Ne-v1-CH-1} can serve as a model for describing unidirectional propagation of waves on shallow water in 1993 \cite{Camassa-PRL-1993}, and pointed out the CH equation is a completely integrable equation and manifested its solitary waves are peakons if $\omega=0$ \cite{Camassa-PRL-1993,Camassa-AAM-1994}. Especially, the CH equation is the earliest integrable models discovered to possess peakon solution different from the soliton solution, and the first-order derivative of this non-smooth peakon solution at the wave peak does not exist. The peakon solution is also considered as a unique type of non-smooth soliton solution to the integrable CH equation \cite{Camassa-PRL-1993}. From Ref. \cite{Camassa-PRL-1993}, the CH equation \eqref{Ne-v1-CH-1} with $\omega=0$ possesses the following Lax pairs
\begin{align}\label{Ne-v1-CH-2}
&f_{\rm{Lp}}:\,\left\{\begin{aligned}
&\phi_{xx}=\bigg[\frac14+\lambda(u-u_{xx})\bigg]\phi\\
&\phi_t=\bigg(\frac{1}{2\lambda}-u\bigg)\phi_x+\frac12u_x\phi
\end{aligned}\right.,
\end{align}
one can deduce the CH equation \eqref{Ne-v1-CH-1} with $\omega=0$ by employing the Lax pairs \eqref{Ne-v1-CH-2} and compatibility condition equation $\phi_{xxt}-\phi_{txx}=0$. We set spectral function $\phi(x,t)\in\mathbb{R}^{1\times 1}$ and spectral parameter $\lambda\in\mathbb{R}$, while spectral parameter $\lambda$ is initialized to $\lambda=1.5$ as well as we initialize $\phi=0$ and make it satisfy the free initial-boundary conditions, then we consider the initial and boundary condition of CH equation \eqref{Ne-v1-CH-1} with $\omega=0$ in spatiotemporal region $[-5,5]\times[0,3]$, shown as bellow
\begin{align}\label{Ne-v1-CH-3}
\begin{split}
&u(x,t=0)=0.9\mathrm{e}^{-|x|},\,x\in[-5,5],\\
&u(-5,t)=0.9\mathrm{e}^{-|-5-0.9t|},\,u(5,t)=0.9\mathrm{e}^{-|5-0.9t|},\,t\in[0,3].
\end{split}
\end{align}

We randomly select $N_{\rm ib}=400$ points based on the aforementioned initial and boundary conditions \eqref{Ne-v1-CH-3}, and extract $N_{\rm c}=10000$ collocation points in residual spatiotemporal region, then we produce the data-set for the LPNN-v1 pertaining to the CH equation. After that, by utilizing LPNN-v1 with training data-set, we successfully and efficiently obtained data-driven peakon solution, as well as corresponding spectral parameter and spectral functions. The relative $L^2$ norm error of the LPNN-v1 model achieves 6.878529$\rm e$-02 for data-driven peakon solution $u(x,t)$ in 148.1191 seconds, and the number of iterations is 530. The spectral parameter learned from LPNN-v1 is $\lambda=1.4812$.

\begin{figure*}[!tb]
\centering
\subfigure[]{\includegraphics[height=1.1in,width=2.2in]{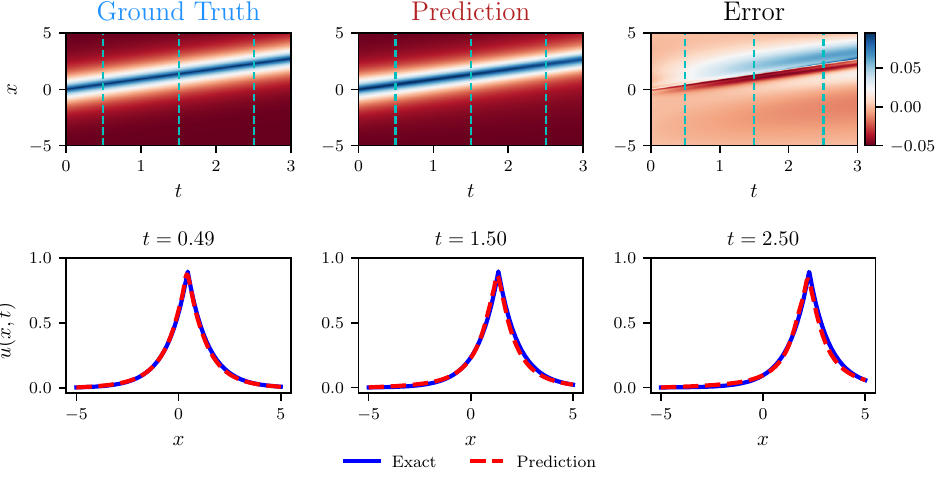}}\hspace{0.15cm}
\subfigure[]{\includegraphics[height=1.1in,width=2.6in]{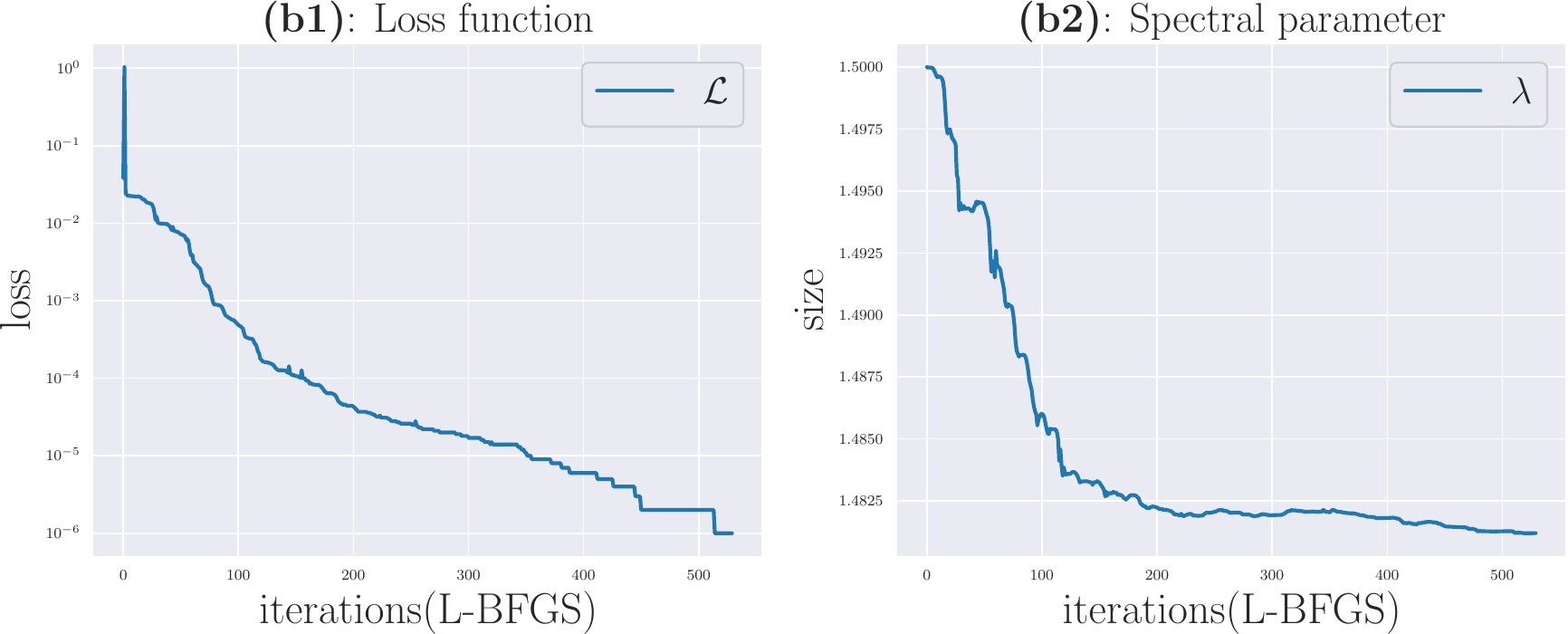}}\\
\subfigure[]{\includegraphics[height=1.4in,width=3.4in]{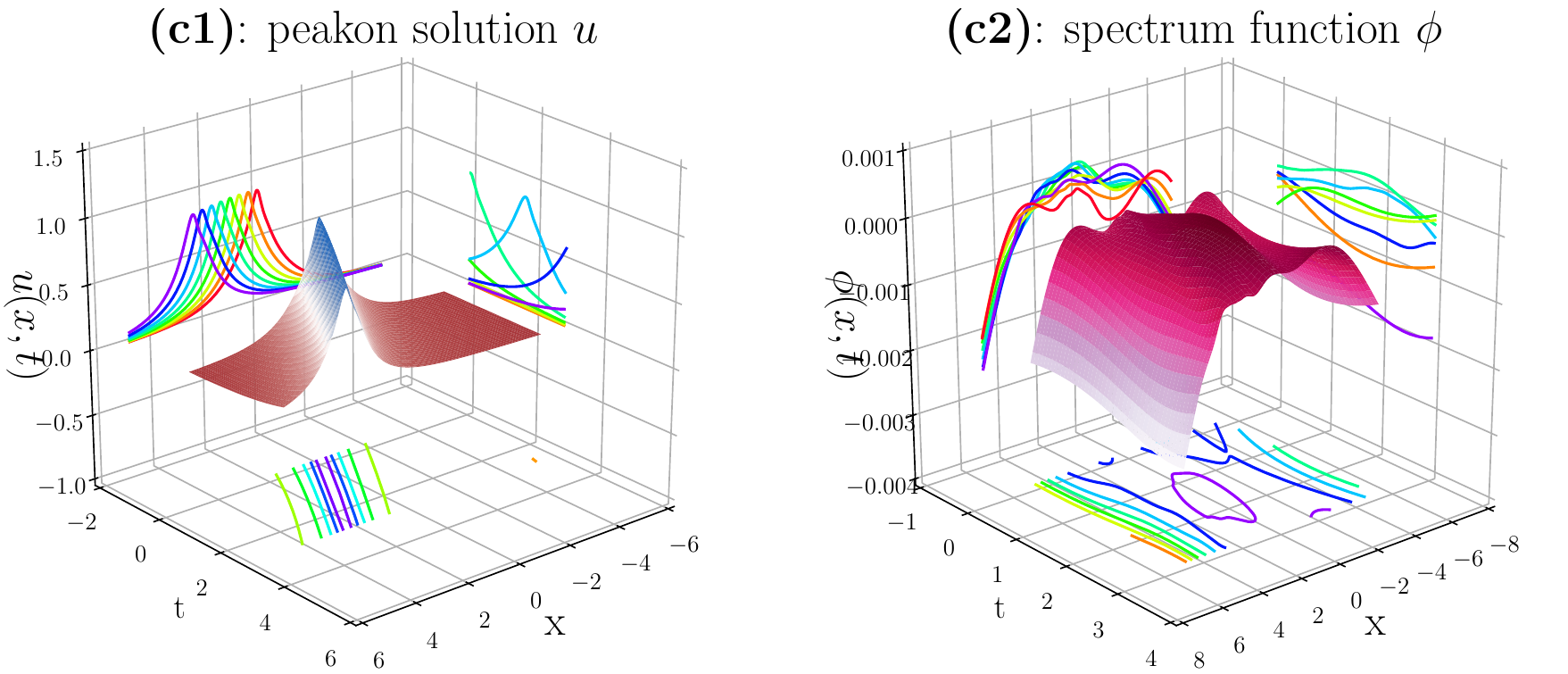}}
\caption{(Color online) The training results of peakon solution $u(x,t)$ and spectral function $\phi(x,t)$ for CH equation arising from the LPNN-v1. (a) The ground truth, prediction and error dynamics density plots, as well as sectional drawings which contain the true and prediction peakon solution at three distinct moments $t=0.49, 1.5, 2.5$; (b) Evolution graphs of the loss function [panel b1] and spectral parameter $\lambda$ [panel b2] in LPNN-v1; (c) The three-dimensional plot with contour map for the data-driven peakon solution [panel c1] and the data-driven spectral function [panel c2] corresponding to spectral parameter $\lambda=1.4812$. \label{figLPNN-v1-CH}}
\end{figure*}

Fig. \ref{figLPNN-v1-CH} exhibits the deep learning results of the data-driven peakon solution $u(x,t)$ and spectral function $\phi(x,t)$ stemming from the LPNN-v1 for the CH equation \eqref{Ne-v1-CH-1} with $\omega=0$. Fig. \ref{figLPNN-v1-CH}(a) showcases the density plots of the ground truth dynamics, prediction dynamics and error dynamics, then give its corresponding amplitude scale size on the right side of density plots, and exhibit the sectional drawings which contain the learned and true solution at three different moments. The evolution curve figures of the loss function [panel b1] and spectral parameter $\lambda$ [panel b2] resulting from the LPNN-v1 are displayed in Fig. \ref{figLPNN-v1-CH}(b). Figs. \ref{figLPNN-v1-CH}(c1) and (c2) indicate respectively the three-dimensional plot with contour map on three planes for the predicted peakon solution and learned spectral function corresponding to spectral parameter $\lambda=1.4812$.

Similarly, in order to further demonstrate the efficiency of LPNN-v1, we compared the training performance with the standard PINN and provided detailed comparison results in Tab. \ref{TabLPNN-v1-CH}. By the way, the abundant data-driven peakon and periodic peakon solutions of the CH equation have been investigated by utilizing the PINN algorithm, interested readers refer to \cite{WangL-PD-2021}. Unlike case 1, in this example we have no fixed spectral parameter $\lambda$ for training, but instead train through LPNN-v1 after giving the initial spectral parameter value. From Tab. \ref{TabLPNN-v1-CH}, we can clearly see that the training error of LPNN-v1 is not significantly improved compared to the standard PINN algorithm, but the training time is more than 5 times faster than the standard PINN algorithm. In contrast with the numerical result of KdV equation in case 1, the training of CH equation in this case is significantly more efficient, it benefits from the complexity for the Lax pairs of the CH equation and KdV equation is not significantly different, but the form of the CH equation itself is more complex than that of the KdV equation. Additionally, the peakon solution in this case is a non-smooth solution, indicating that LPNN-v1 is also capable of efficiently recovering non-smooth solution.

\begin{table}[htbp]
  \caption{Performance comparison between LPNN-v1 and conventional PINN for solving CH equation}
  \label{TabLPNN-v1-CH}
  \centering
  \scalebox{0.8}{
  \begin{tabular}{l|c|c|c|c|c|c}
  \toprule
  \small{\textbf{Types}} & $\bm{\mathrm{x}}\times t$ & $\mathcal{D}_{\mathrm{ib}},\mathcal{D}_{\mathrm{c}}$ & optimizer & $\lambda$ & $L^2$ norm error & training time \\
  \hline
  PINN   & [-5,5]$\times$[0,3] & 400,10000 & L-BFGS & $\backslash$ & 6.882064$\rm e$-02 & 819.0868s \\
  \hline
  LPNN-v1 & [-5,5]$\times$[0,3] & 400,10000 & L-BFGS & 1.4812 & 6.878529$\rm e$-02 & 148.1191s \\
  \bottomrule
  \end{tabular}}
\end{table}

$\bullet$ \textbf{Case 3: Kadomtsev-Petviashvili equation}

In the first two examples, we efficiently solved two classical (1+1)-dimensional integrable models using LPNN-v1. The (2+1)-dimensional integrable systems usually have three independent variables $\{x, y, t\}$, where $x$ and $y$ usually refer to space variables and $t$ refers to time variable. To further verify the performance of LPNN-v1 in high-dimensional integrable systems, we consider a typical (2+1)-dimensional integrable PDE in this example, namely the KP equation \cite{Kadomtsev-DAN-1970}. The KP equation was derived by Kadomtsev and Petviashvili to examine the stability of the single-soliton of the KdV equation under transverse perturbations in 1970, and it is relevant for most applications in which the KdV equation arises. The KP equation can be written as
\begin{align}\label{Ne-v1-KP-1}
(u_t+6uu_x+u_{xxx})_x+3\sigma^2u_{yy}=0,
\end{align}
where $\sigma^2=-1$ or $\sigma^2=1$. The Eq. \eqref{Ne-v1-KP-1} is called the KPI equation if $\sigma^2=-1$, and the KPII equation if $\sigma^2=1$. Since the KP equation \eqref{Ne-v1-KP-1} degenerates into the one-dimensional KdV equation \eqref{Ne-v1-KdV-1} as $u$ is independent of $y$, the KP equation is the natural generalization of the KdV equation. The KP equation describes the motion of two dimensional water wave and possesses important applications in fluid mechanics and theoretical physics.

Correspondingly, we can directly write the Lax pairs spectral problem of the KP equation \eqref{Ne-v1-KP-1} \cite{Novikov-M-1984}, as shown in the following equation
\begin{align}\label{Ne-v1-KP-2}
&f_{\rm{Lp}}:\,\left\{\begin{aligned}
&\phi_{y}=-\sigma^{-1}\phi_{xx}-\sigma^{-1}u\phi\\
&\phi_t=-4\phi_{xxx}-6u\phi_x-(3u_x-3\sigma\partial^{-1}_xu_y)\phi
\end{aligned}\right..
\end{align}
We can directly derive the KP equation \eqref{Ne-v1-KP-1} by utilizing the compatibility condition equation $\phi_{yt}-\phi_{ty}=0$. Due to the complexity and peculiarity of high-dimensional integrable systems, Lax pairs of some high-dimensional integrable systems [including the KP equation \eqref{Ne-v1-KP-1} in this part and high-dimensional KdV equation \eqref{Ne-v1-HDKdV-1} in next example] do not involve corresponding spectral parameter $\lambda$ and only contain spectral function $\phi(x,y,t)$ and potential function $u(x,y,t)$. Specifically, we consider the KPII equation with taking $\sigma=1$, then study the following initial and boundary conditions
\begin{align}\label{Ne-v1-KP-3}
\begin{split}
&u(x,y,t=-0.5)=\frac{18\mathrm{e}^{3x+6y+31.5}}{(1+\mathrm{e}^{3x+6y+31.5})^2},\,\bm{\mathrm{x}}\in\Omega,\\
&u(x,y,t)=\frac{18\mathrm{e}^{3x+6y-63t}}{(1+\mathrm{e}^{3x+6y-63t})^2},\,\bm{\mathrm{x}}\in\partial\Omega,\,t\in[-0.5,0.5].
\end{split}
\end{align}
here $\bm{\mathrm{x}}=\{x,y\},\,\Omega=[-3,3]\times[-3,3]$. The spectral function $\phi\in\mathbb{R}^{1\times 1}$ satisfied free initial-boundary condition and initialized to $\phi(x,y,t)=0$. We select initial and boundary points $N_{\rm ib}=500$ from the initial-boundary conditions \eqref{Ne-v1-KP-3}, and collocation points $N_{\rm c}=10000$. After 1125 L-BFGS optimization in LPNN-v1, the relative $L^2$ norm error for line-soliton solution $u(x,y,t)$ reaches 5.867296$\rm e$-03 in 2767.3306 seconds. Generally, if the region where $u$ is far from zero forms a band on the $(x, y)$ plane, this kind of solutions are called “line-solitons”. This does not happen in (1+1)-dimensional integrable systems.

\begin{figure*}[!htbp]
\centering
\subfigure[]{\includegraphics[height=1.0in,width=1.6in]{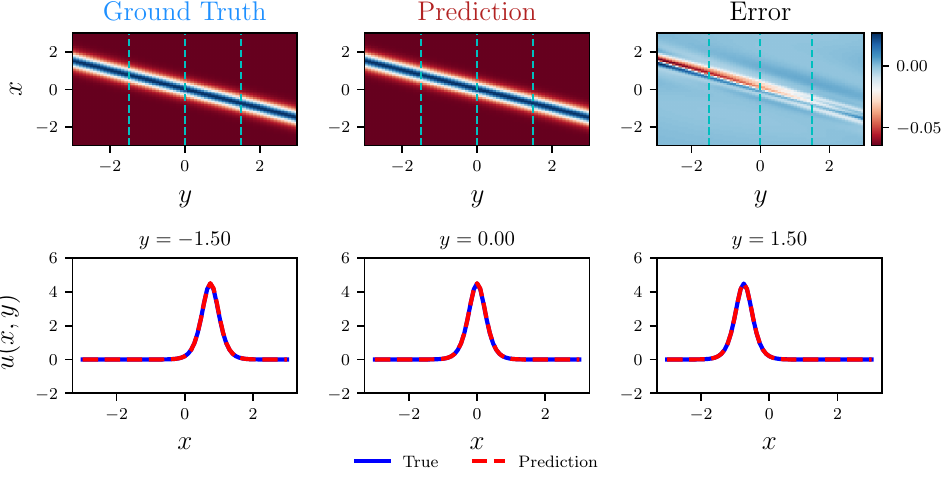}}\hspace{0.15cm}
\subfigure[]{\includegraphics[height=1.0in,width=1.4in]{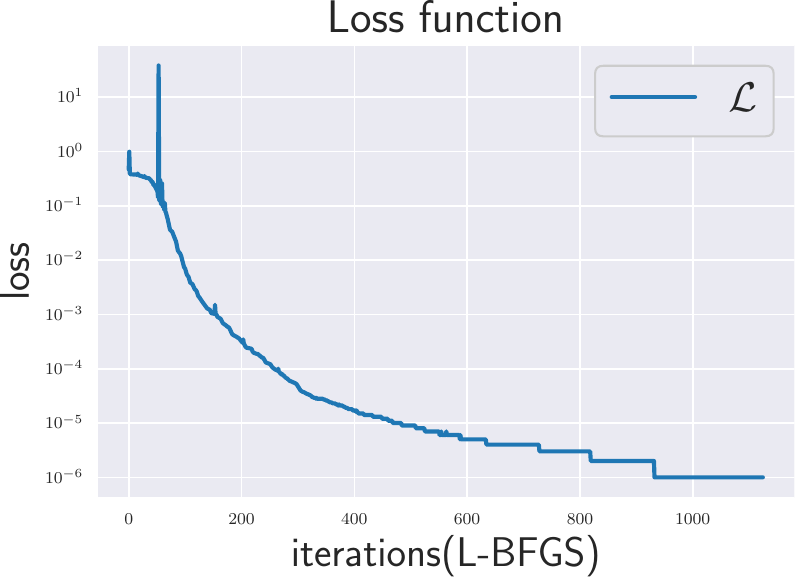}}\hspace{0.15cm}
\subfigure[]{\includegraphics[height=1.0in,width=1.2in]{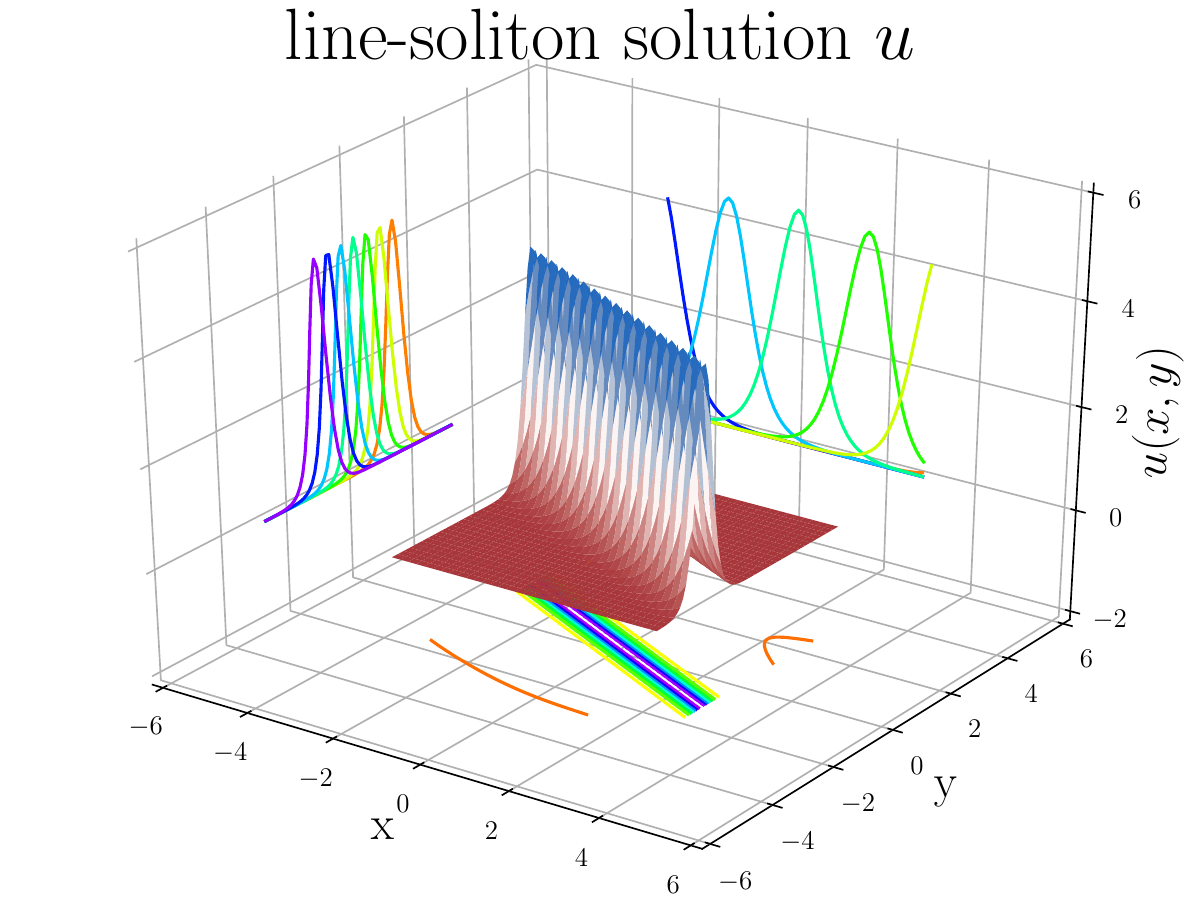}}
\subfigure[]{\includegraphics[height=1.0in,width=1.2in]{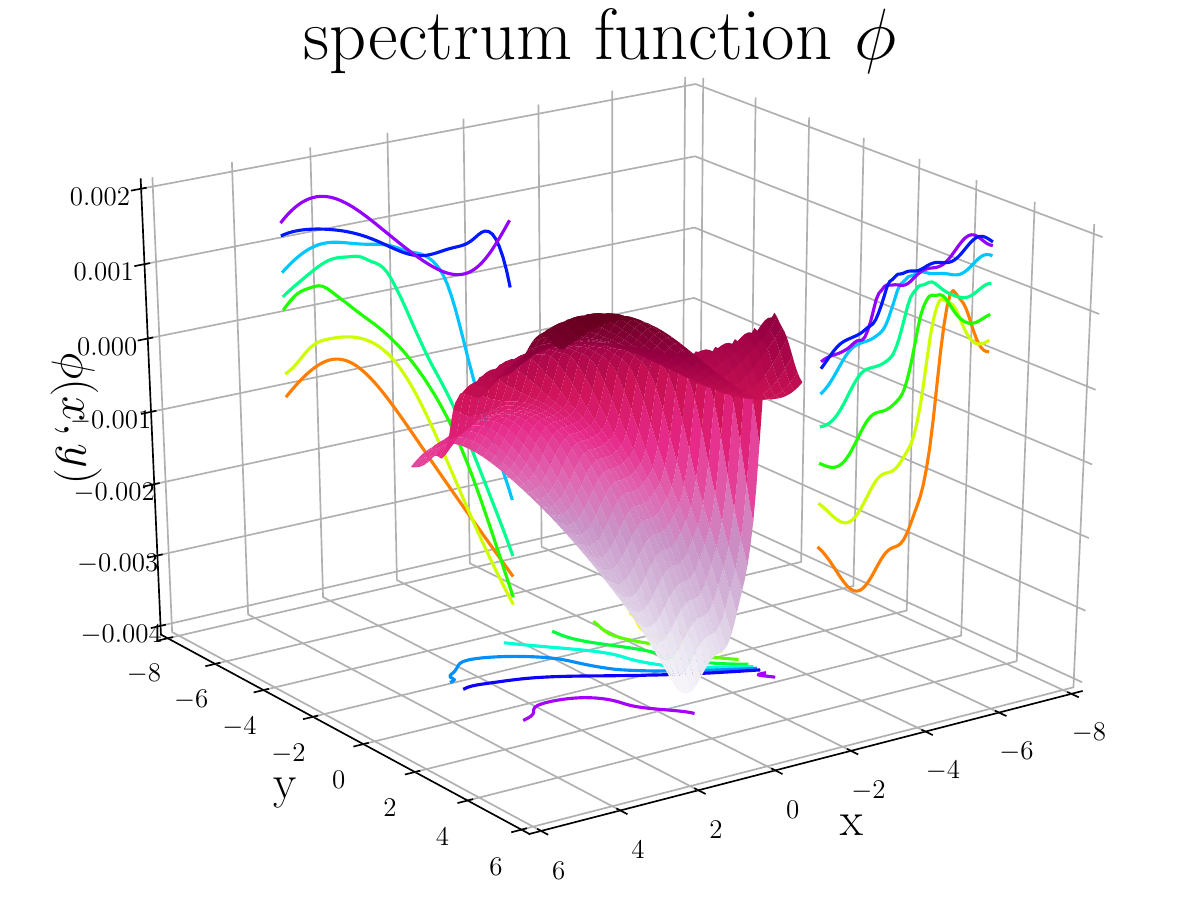}}
\caption{(Color online) The training results of line-soliton solution $u(x,y,t)$ and spectral function $\phi(x,y,t)$ at time $t=0$ for KP equation arising from the LPNN-v1. (a) The ground truth, prediction and error dynamics density plots, as well as sectional drawings which contain the true and prediction line-soliton solution at three distinct $y$-dots $y=-1.5, 0, 1.5$; (b) Evolution graphs of the loss function in LPNN-v1; (c) The three-dimensional plot with contour map for the data-driven line-soliton solution; (d) The three-dimensional plot with contour map for the data-driven spectral function. \label{figLPNN-v1-KP}}
\end{figure*}

Fig. \ref{figLPNN-v1-KP} displays the numerical results obtained by solving the KP equation by applying LPNN-v1. Fig. \ref{figLPNN-v1-KP}(a) showcases three different density plots and sectional drawings at three distinct $y$-dots. In the process of LPNN-v1 optimization training using L-BFGS, the evolution figures of loss function [panel b1] and spectral parameter [panel b2] are given in Fig. \ref{figLPNN-v1-KP}(b). Figs. \ref{figLPNN-v1-KP}(c) and Fig. \ref{figLPNN-v1-KP}(d) exhibit the three-dimensional plots of the line-soliton solution $u(x,y,t)$ and spectral function $\phi(x,y,t)$, respectively. Tab. \ref{TabLPNN-v1-KP} provides a detailed results comparison of using PINN and LPNN-v1 to solve the line-soliton solution of the KP equation. Surprisingly, comparing with the PINN, one can find that the LPNN-v1 not only reduces training time but also improves training accuracy by about twice. The numerical results indicate that LPNN-v1 is also suitable for efficiently solving high-dimensional integrable systems and their Lax pairs spectral problems.

\begin{table}[htbp]
  \caption{Performance comparison between LPNN-v1 and conventional PINN for solving KP equation}
  \label{TabLPNN-v1-KP}
  \centering
  \scalebox{0.8}{
  \begin{tabular}{l|c|c|c|c|c|c}
  \toprule
  \small{\textbf{Types}} & $\bm{\mathrm{x}}\times t$ & $\mathcal{D}_{\mathrm{ib}},\mathcal{D}_{\mathrm{c}}$ & optimizer & $L^2$ norm error & training time \\
  \hline
  PINN   & [-3,3]$\times$[-3,3]$\times$[-0.5,0.5] & 500,10000 & L-BFGS & 1.037711$\rm e$-02 & 4757.5630s \\
  \hline
  LPNN-v1 & [-3,3]$\times$[-3,3]$\times$[-0.5,0.5] & 500,10000 & L-BFGS & 5.867296$\rm e$-03 & 2767.3306s \\
  \bottomrule
  \end{tabular}}
\end{table}

$\bullet$ \textbf{Case 4: High-dimensional Korteweg-de Vries equation}

In this part, we  further consider the following high-dimensional KdV equation
\begin{align}\label{Ne-v1-HDKdV-1}
u_{ty}+u_{xxxy}+3(u_yu_x)_x-u_{xx}+2u_{yy}=0,
\end{align}
here $\bm{\mathrm{x}}=\{x,y\}\in\Omega$ indicates the $2$-dimensional space in Eq. \eqref{M1a}. Wazwaz firstly put forward the new (2+1)-dimensional KdV equation, and pointed out the integrability of the new (2+1)-dimensional KdV equation is investigated via using the Painlev\'{e} test \cite{Wazwaz-NPB-2020}. Afterwards, with the aid of Bell polynomials theory, we derived the bilinear formalism, bilinear B\"{a}cklund transformations and Lax pairs of the (2+1)-dimensional KdV equation \eqref{Ne-v1-HDKdV-1}, and obtained the $N$-soliton solution base on the bilinear formalism, the lump solution and quasiperiodic wave solutions along with their asymptotic properties in Ref. \cite{Pu-AMAS-2022}. Therefore, we can directly write the Lax pairs of operator form \eqref{M1b} for the high-dimensional KdV equation \eqref{Ne-v1-HDKdV-1}, as shown below
\begin{align}\label{Ne-v1-HDKdV-2}
&f_{\rm{Lp}}:\,\left\{\begin{aligned}
&\phi_{xy}=\frac13\phi-\phi u_y\\
&\phi_t=-3u_x\phi_x-\phi_{xxx}-2\phi_y
\end{aligned}\right..
\end{align}
We can derive the high-dimensional KdV equation \eqref{Ne-v1-HDKdV-1} via the compatibility condition equation $\phi_{xyt}=\phi_{txy}$. Now we let spectral function $\phi(x,y,t)\in\mathbb{R}^{1\times 1}$ satisfied free initial-boundary condition and initialized to $\phi(x,y,t)=0$ in spatiotemporal region $[-10,10]\times[-10,10]\times[-5,5]$, namely $\Omega=[-10,10]\times[-10,10]$. Then we consider the following initial and boundary conditions of high-dimensional KdV equation \eqref{Ne-v1-HDKdV-1}
\begin{align}\label{Ne-v1-HDKdV-3}
\begin{split}
&u(x,y,t=-5)=\frac{15+2x+2y}{(0.5x+y+10)^2+(-2.5+0.5x)^2+3},\,\bm{\mathrm{x}}\in\Omega,\\
&u(x,y,t)=\frac{-3t+2x+2y}{(0.5x+y-2t)^2+(0.5t+0.5x)^2+3},\,\bm{\mathrm{x}}\in\partial\Omega,\,t\in[-5,5].
\end{split}
\end{align}
Owing to the $2$-dimensional space of high-dimensional KdV equation \eqref{Ne-v1-HDKdV-1}, we can obtain the $\partial\Omega$ has four boundary surfaces. Then we use 400 initial and boundary points, as well as 20000 collocation points in LPNN-v1, and obtain the 9.171556$\rm e$-03 relative $L^2$ norm error for data-driven lump solution $u(x,y,t)$ and spectral function $\phi(x,y,t)$ of Lax pairs spectral problem \eqref{Ne-v1-HDKdV-2} by using 641 L-BFGS optimization in 963.9654 seconds. The lump solution is a special type of rational function solution in high-dimensional integrable systems, which is localized in all directions of space, it appears in various fields such as Bose-Einstein condensates, optics and marine science \cite{Satsuma-JMP-1979}.

\begin{figure*}[!htbp]
\centering
\subfigure[]{\includegraphics[height=1.0in,width=1.6in]{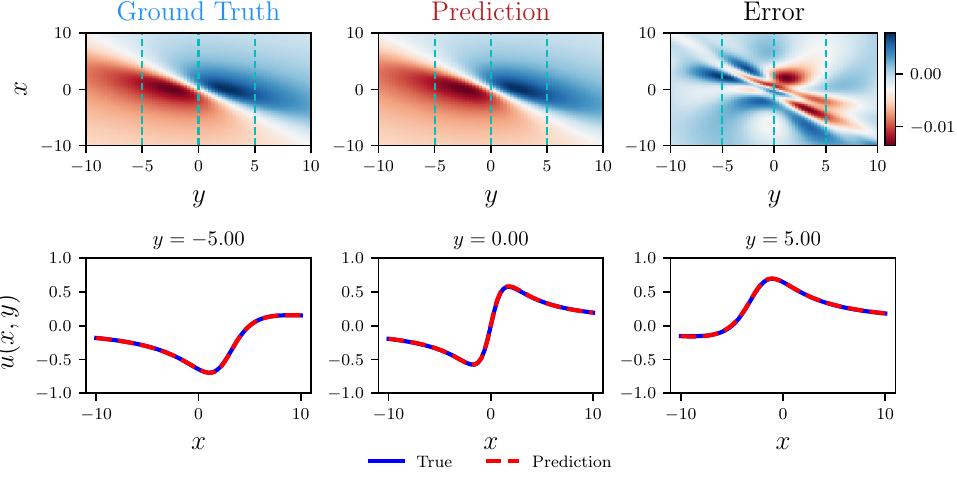}}\hspace{0.15cm}
\subfigure[]{\includegraphics[height=1.0in,width=1.4in]{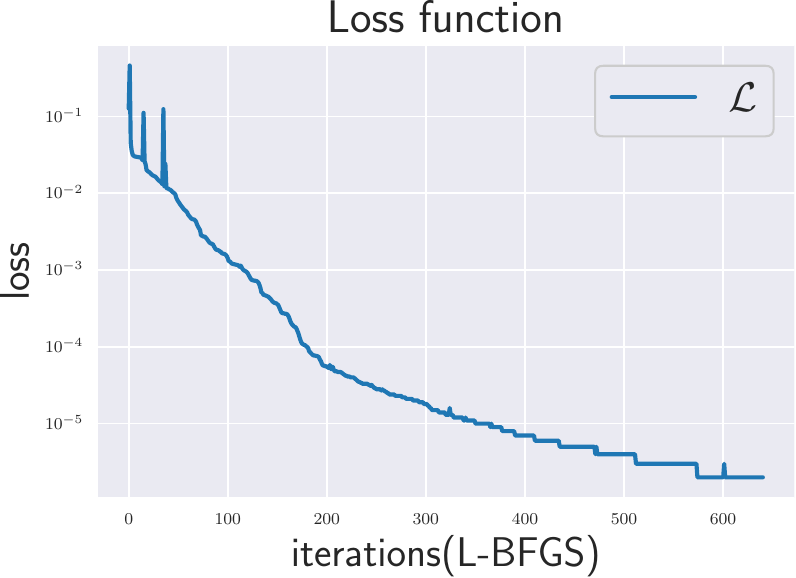}}\hspace{0.15cm}
\subfigure[]{\includegraphics[height=1.0in,width=1.2in]{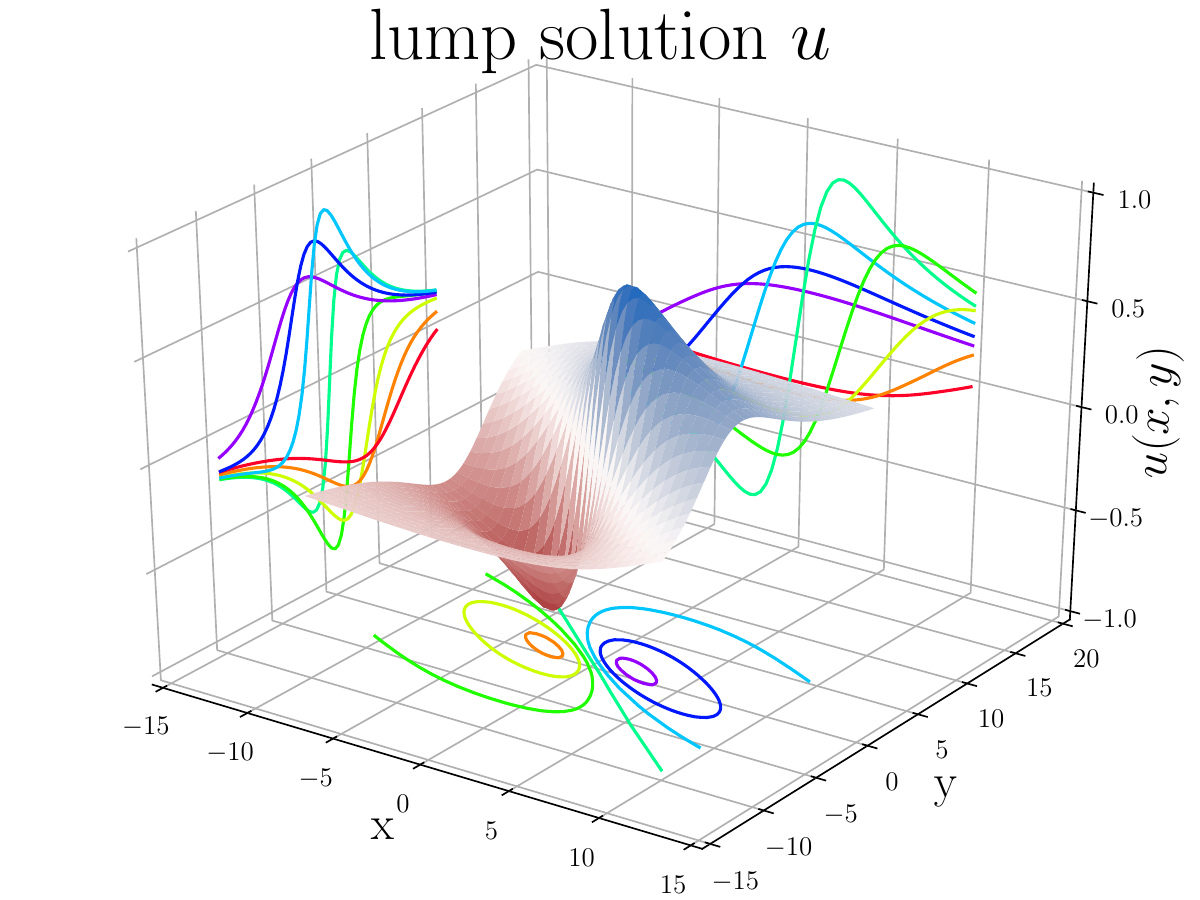}}
\subfigure[]{\includegraphics[height=1.0in,width=1.2in]{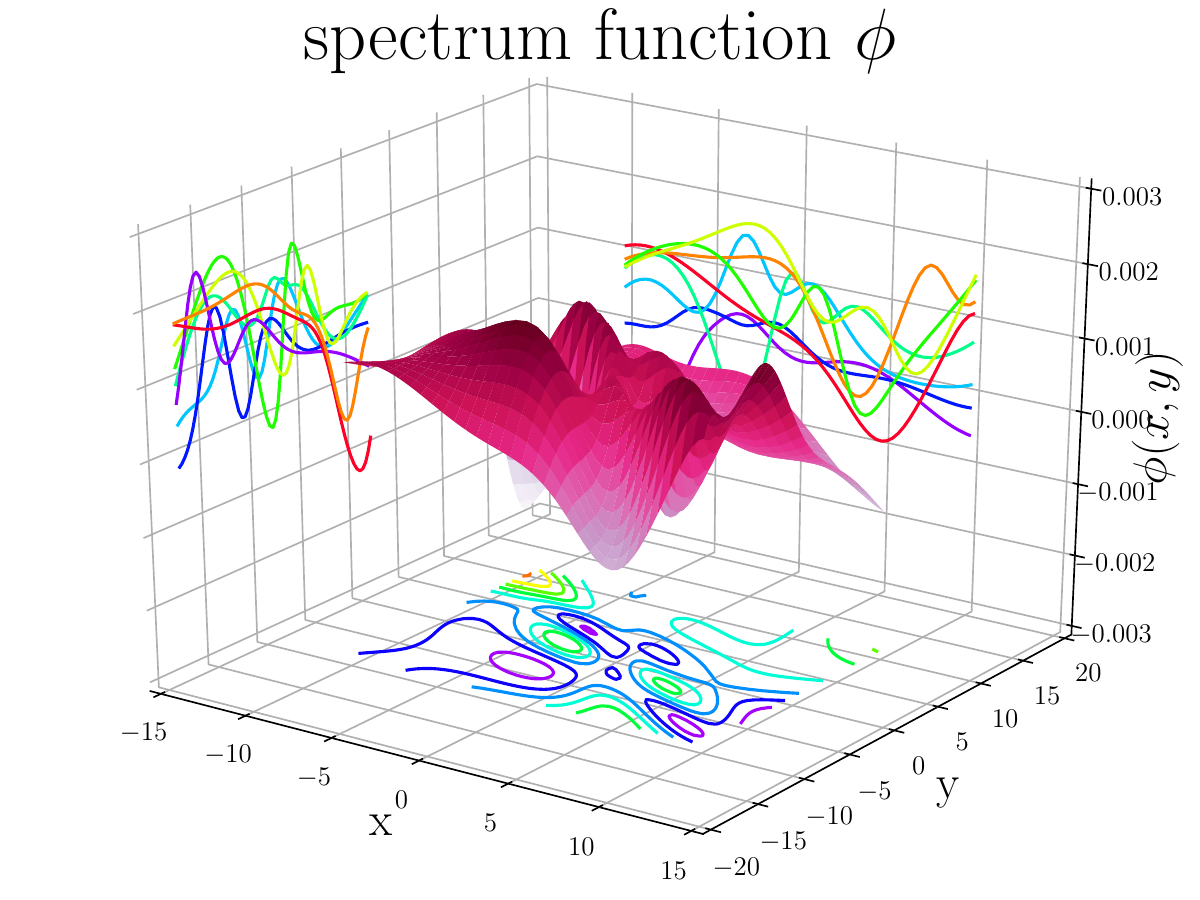}}
\caption{(Color online) The training results of lump solution $u(x,y,t)$ and spectral function $\phi(x,y,t)$ at time $t=0$ for high-dimensional KdV equation arising from the LPNN-v1. (a) The ground truth, prediction and error dynamics density plots, as well as sectional drawings which contain the true and prediction lump solution $u(x,y)$ at three distinct $y$-dots $y=-5, 0, 5$; (b) Evolution graph of the loss function in LPNN-v1; (c) The three-dimensional plot with contour map for the data-driven lump solution $u(x,y)$; (d) The three-dimensional plot with contour map for the data-driven spectral function $\phi(x,y)$. \label{figLPNN-v1-HDKdV}}
\end{figure*}

Fig. \ref{figLPNN-v1-HDKdV} depicts the deep learning results of the data-driven lump solution $u(x,y,t)$ and spectral function $\phi(x,y,t)$ at time $t=0$ stemming from the LPNN-v1 for
the high-dimensional KdV equation \eqref{Ne-v1-HDKdV-1}. Specifically, as shown in Fig. \ref{figLPNN-v1-HDKdV}(a), we display the density plots of the true dynamics, prediction dynamics and error dynamics, then showcase its corresponding amplitude scale size on the right side of density plots, and exhibit the sectional drawings which contain the learned and true solution at three different $y$-dots. The evolution curve figure of the loss function arising from the LPNN-v1 is displayed in Fig. \ref{figLPNN-v1-HDKdV}(b). Figs. \ref{figLPNN-v1-HDKdV}(c) and (d) indicate respectively the three-dimensional plot with contour map on three planes for the predicted lump solution $u(x,y)$ and learned spectral function $\phi(x,y)$. Generally, we can also fix the $x$-axis [or $y$-axis] to visualize the relevant deep learning results for data-driven lump solution and spectral function in the $(y, t)$-dimension [or $(x, t)$-dimension].

Under the same training conditions, we also used traditional PINN to solve the high-dimensional KdV equation, and presented a comparison of the efficiency and accuracy of the two algorithms in Tab. \ref{TabLPNN-v1-HDKdV}. As shown in Tab. \ref{TabLPNN-v1-HDKdV}, compared with PINN, we can know that the LPNN-v1 decreased training time by more than twice at the cost of sacrificing training accuracy. The numerical results further demonstrate that LPNN-v1 is equally effective for solving high-dimensional integrable systems.

\begin{table}[htbp]
  \caption{Performance comparison between LPNN-v1 and conventional PINN for solving high-dimensional KdV equation}
  \label{TabLPNN-v1-HDKdV}
  \centering
  \scalebox{0.8}{
  \begin{tabular}{l|c|c|c|c|c}
  \toprule
  \small{\textbf{Types}} & $\bm{\mathrm{x}}\times t$ & $\mathcal{D}_{\mathrm{ib}},\mathcal{D}_{\mathrm{c}}$ & optimizer & $L^2$ norm error & training time \\
  \hline
  PINN   & [-10,10]$\times$[-10,10]$\times$[-5,5] & 400,20000 & L-BFGS & 5.416409$\rm e$-03 & 2002.7561s \\
  \hline
  LPNN-v1 & [-10,10]$\times$[-10,10]$\times$[-5,5] & 400,20000 & L-BFGS & 9.171556$\rm e$-03 & 963.9654s \\
  \bottomrule
  \end{tabular}}
\end{table}

In this subsection, we efficiently solved the KdV equation, CH equation, KP equation and high-dimensional KdV equation using LPNN-v1 with a novel network framework. In addition to obtaining data-driven soliton solution, non-smooth peakon solution, line-soliton solution and lump solution, we also learned and obtained the corresponding spectral parameter and spectral function of Lax pairs spectral problems for the first time. Due to the fact that these integrable systems have simpler Lax pairs than these equations itself, the Lax pairs informed part of LPNN-v1 is simpler and more direct than the PDE informed of standard PINN, thus efficiently solving integrable systems and Lax pairs spectral problems. Surprisingly, our proposed LPNN-v1 significantly improves the training efficiency in solving non-smooth solution and studying high-dimensional integrable system, with training accuracy is similar to or even higher than the standard PINN. Usually, high-dimensional integrable systems themselves have a relatively complex form, while their Lax pairs have a simpler form compared to PDE itself, hence the LPNN-v1 is particularly suitable for studying high-dimensional integrable systems.

\subsection{high-accuracy solving performance of LPNN-v2}
In the following subsections, we delve into the data-driven localized wave solutions and spectral problems for other significant and classical integrable systems with Lax pairs by applying LPNN-v2, and provide comprehensive numerical results along with kinetic analysis.

$\bullet$ \textbf{Case 1: Sine-Gordon equation}

The SG equation is classical integrable model, which is introduced in 1939 by Frenkel and Kontorova as a model for the dynamics of crystal dislocations \cite{Frenkel-JP-1939}. The SG equation has found a variety of applications, including Bloch wall dynamics in ferromagnetics and ferroelectrics, fluxon propagation in long Josephson (superconducting) junctions, self-induced transparency in nonlinear optics, spin waves in the A-phase of liquid $^3$He at temperatures near to 2.6 mK, and a simple, one-dimensional model for elementary particles \cite{Scott-OUP-2003}. The SG equation and its Lax pairs [matrix form] can be represented as

\begin{align}\label{Ne-v2-SG-1}
&u_{xt}=\sin(u),
\end{align}
\begin{align}\label{Ne-v2-SG-2}
f_{\rm{Lp}}:\,\bigg\{\begin{aligned}
\Phi_{x}=M\Phi\\
\Phi_t=N\Phi
\end{aligned},\,
M=\begin{bmatrix} -\mathrm{i}\lambda & -\frac12u_x  \\ \frac12u_x & \mathrm{i}\lambda \end{bmatrix},\,N=\begin{bmatrix} \frac{\mathrm{i}}{4\lambda}\cos(u) & \frac{\mathrm{i}}{4\lambda}\sin(u)  \\ \frac{\mathrm{i}}{4\lambda}\sin(u) & -\frac{\mathrm{i}}{4\lambda}\cos(u) \end{bmatrix},
\end{align}
here ``$\mathrm{i}$'' indicates imaginary unit, and we have $u(x,t)\in\mathbb{R}^{1\times 1}$, spectral function $\Phi(x,t)\in\mathbb{C}^{2\times 1}$, spectral parameter $\lambda=\lambda_1+\lambda_2\mathrm{i}\in\mathbb{C}$. Thus we set $\Phi(x,t)=(\phi_1(x,t),\phi_2(x,t))^{\rm T}$ and $\phi_1(x,t)=\phi_{11}(x,t)+\mathrm{i}\phi_{12}(x,t),\phi_2(x,t)=\phi_{21}(x,t)+\mathrm{i}\phi_{22}(x,t)$, in which $\phi_{ij}(x,t)\in\mathbb{R}^{1\times 1}[i,j=1,2]$. Accordingly, we can derive the SG equation \eqref{Ne-v2-SG-1} through the zero curvature equation \eqref{I5}, we let $\phi_{ij}$ satisfy free initial-boundary conditions and they are initialized to $\phi_{ij}=0$, and we initialize spectral parameter $\lambda$ to $0+0\mathrm{i}$. Then we consider the initial and boundary conditions of SG equation in spatiotemporal region $[-5,5]\times[-5,5]$ as bellow
\begin{align}\label{Ne-v2-SG-3}
\begin{split}
&u(x,t=-5)=4\arctan(\mathrm{e}^{x-5}),\,x\in[-5,5],\\
&u(-5,t)=4\arctan(\mathrm{e}^{-5+t}),\,u(5,t)=4\arctan(\mathrm{e}^{5+t}),\,t\in[-5,5].
\end{split}
\end{align}

We use 500 initial and boundary points, as well as 10000 collocation points in LPNN-v2, then obtain the 1.962527$\rm e$-04 relative $L^2$ norm error for data-driven
kink solution $u(x,t)$ by using 666 L-BFGS optimization in 215.7550 seconds. Moreover, we also numerically learned the spectral parameter $\lambda=-0.000402+0.000416\mathrm{i}$ and their corresponding spectral function $\Phi(x,t)=(\phi_1(x,t),\phi_2(x,t))^{\rm T}$ in spectral problem \eqref{Ne-v2-SG-2}.

\begin{figure*}[!tb]
\centering
\subfigure[]{\includegraphics[height=1.0in,width=2.0in]{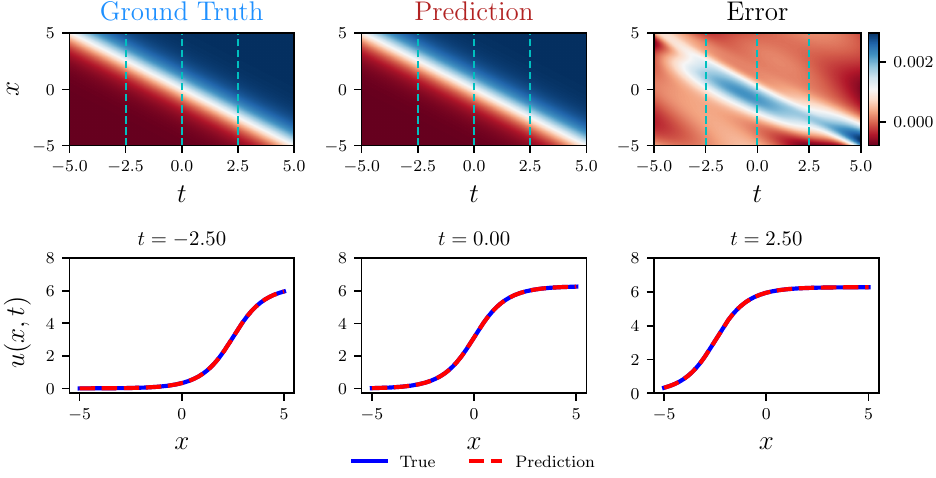}}\hspace{0.5cm}
\subfigure[]{\includegraphics[height=1.0in,width=3.6in]{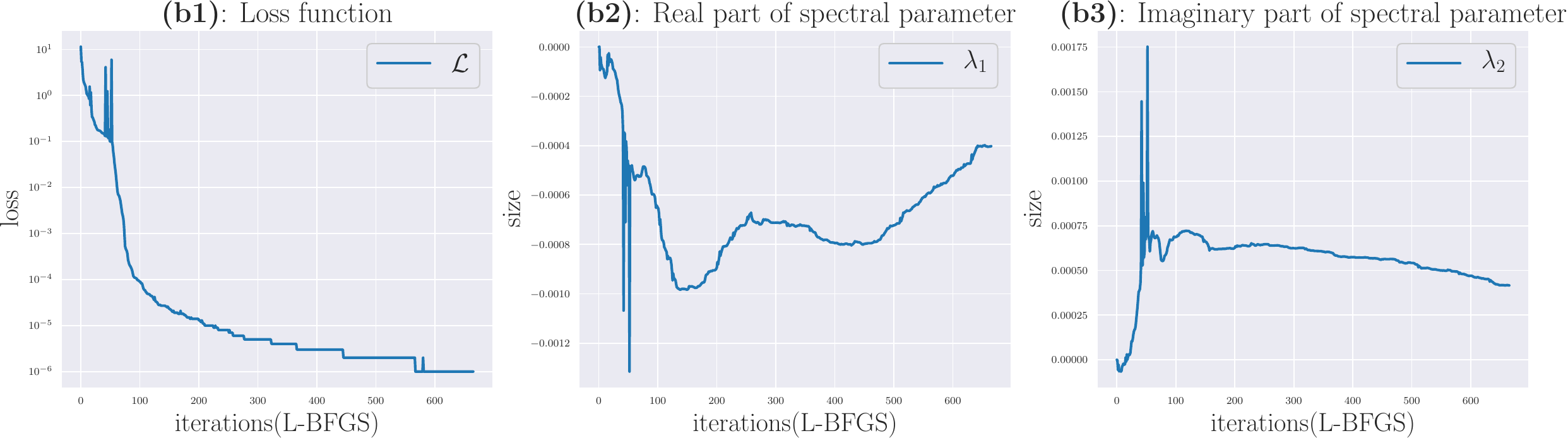}}\\
\subfigure[]{\includegraphics[height=1.4in,width=5.8in]{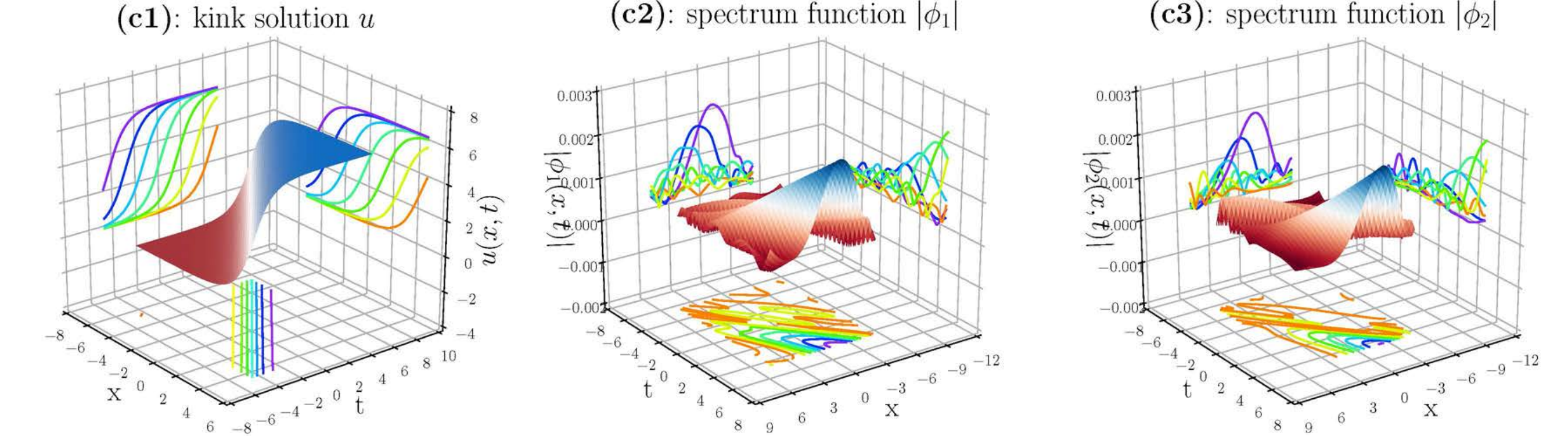}}
\caption{(Color online) The training results of kink solution $u(x,t)$ and spectral functions $\phi_1(x,t),\phi_2(x,t)$ for SG equation arising from the LPNN-v2. (a) The ground truth, prediction and error dynamics density plots, as well as sectional drawings which contain the true and prediction kink solution at three distinct moments $t=-2.5, 0, 2.5$; (b) Evolution graphs of the loss function [panel b1] and spectral parameter $\lambda=\lambda_1+\mathrm{i}\lambda_2$ [panels b2 and b3] in LPNN-v2; (c) The three-dimensional plots with contour map for the data-driven kink solution  [panel c1] and spectral functions [panels c2, c3] corresponding to spectral parameter $\lambda=-0.000402+0.000416\mathrm{i}$. \label{figLPNN-v2-SG}}
\end{figure*}

Fig. \ref{figLPNN-v2-SG} displays the data-driven training results of the kink solution $u(x,t)$ and spectral functions $\phi_1(x,t),\phi_2(x,t)$ by utilizing the LPNN-v2 with the initial-boundary value conditions of the SG equation. The upper panel of Fig. \ref{figLPNN-v2-SG}(a) depicts various dynamic density plots, including true, learned dynamics as well as error dynamics with corresponding amplitude scale size on the right side, and the bottom panel of Fig. \ref{figLPNN-v2-SG}(a) presents sectional drawing at different moments. The evolution curve figures for the loss function [panel (b1)] and spectral parameter [panels (b2), (b3)] arising from the LPNN-v2 with L-BFGS are displayed in Fig. \ref{figLPNN-v2-SG}(b). The three-dimensional plots with contour map on three planes for the data-driven kink solution and spectral functions have been displayed in Fig. \ref{figLPNN-v2-SG}(c), in which left panel is three-dimensional figures of kink solution $u(x,t)$, while the middle and right panel are three-dimensional figures of spectral functions $\phi_1(x,t),\phi_2(x,t)$.

Similarly, we exhibit that comparison of relative $L^2$ norm error and training time between LPNN-v2, LPNN-v1 and conventional PINN for solving SG equation in Table \ref{TabLPNN-v2-SG}. From this table, we found that in the context of consistent training parameter information, traditional PINN has the shortest training time, LPNN-v2 has the highest training accuracy, while LPNN-v1 has both time and accuracy at the intermediate level. Owing to the SG equation form is more concise compared to its Lax pairs, so LPNN-v1 cannot complete efficient training, but LPNN-v2 can achieve high training accuracy. Therefore, we can see that LPNN-v2 can achieve high training accuracy, but at the same time it may consume more training time.

\begin{table}[htbp]
  \caption{Performance comparison between LPNN-v2, LPNN-v1 and conventional PINN for solving SG equation}
  \label{TabLPNN-v2-SG}
  \centering
  \scalebox{0.8}{
  \begin{tabular}{l|c|c|c|c|c|c|c}
  \toprule
  \small{\textbf{Types}} & $\bm{\mathrm{x}}\times t$ & $\mathcal{D}_{\mathrm{ib}},\mathcal{D}_{\mathrm{c}}$ & optimizer & $\lambda$ & $L^2$ norm error & training time \\
  \hline
  PINN   & [-5,5]$\times$[-5,5] & 500,10000 & L-BFGS & $\backslash$ & 7.025371$\rm e$-04 & 19.8829s \\
  \hline
  LPNN-v1 & [-5,5]$\times$[-5,5] & 500,10000 & L-BFGS & 0.004767-0.01552$\mathrm{i}$ & 2.711373$\rm e$-03 & 38.0464s \\
  \hline
  LPNN-v2 & [-5,5]$\times$[-5,5] & 500,10000 & L-BFGS & -0.000402+0.000416$\mathrm{i}$ & 1.962527$\rm e$-04 & 215.7550s \\
  \bottomrule
  \end{tabular}}
\end{table}

$\bullet$ \textbf{Case 2: Modified KdV equation}

The mKdV equation, which was derived in the study of an-harmonic lattices, is also one of the most important models in integrable systems and can be regarded as the KdV equation \eqref{Ne-v1-KdV-1} with a cubic nonlinearity \cite{Zabusky-NPDE-1967}. Moreover, the exact solutions of the KdV equation and the mKdV equation can be transformed into each other via the explicit Miura transformation, which provides an important foundation for the development of Miura transformation theory \cite{Fordy-JMP-1980}. We directly present the mKdV equation and its Lax pairs spectral problem, as follows:
\begin{align}\label{Ne-v2-mKdV-1}
u_t+6u^2u_{x}+u_{xxx}=0,
\end{align}
\begin{align}\label{Ne-v2-mKdV-2}
\begin{split}
&f_{\rm{Lp}}:\,\bigg\{\begin{aligned}
\Phi_{x}=M\Phi\\
\Phi_t=N\Phi
\end{aligned},\,
M=\begin{bmatrix} \lambda & u  \\ -u & -\lambda \end{bmatrix},\\
&N=\begin{bmatrix} -4\lambda^3-2u^2\lambda & -4u\lambda^2-2u_x\lambda-2u^3-u_{xx}  \\ 4u\lambda^2-2u_x\lambda+2u^3+u_{xx} & 4\lambda^3+2u^2\lambda \end{bmatrix}.
\end{split}
\end{align}

Different from the KdV equation, the Lax pairs \eqref{Ne-v2-mKdV-2} of the mKdV equation are more complex compared to the mKdV equation itself, so we choose LPNN-v2 to solve and study its localized wave solution and spectral problem. Unlike the previous handling of the SG equation, we take spectral parameter $\lambda\in\mathbb{R}$ and initialize it to $\lambda=1$. Particularly, we select spectral function $\Phi(x,t)\in\mathbb{R}^{2\times 2}$, namely we have second-order matrix $\Phi(x,t)=[\phi_{ij}]_{2\times 2},\, i,j=1,2$. We initialize the spectral functions to $\phi_{ij}=0$ while satisfying the free initial-boundary condition. When the spatiotemporal variables $\{x,t\}\in[-3,3]\times[-1.5,1.5]$, we consider the solution of the mKdV equation with Lax pairs to have the following initial and boundary conditions
\begin{align}\label{Ne-v2-mKdV-3}
\begin{split}
&q(x,t=-0.5)=2\mathrm{sech}(2x+6),\,x\in[-4,4],\\
&q(-3,t)=2\mathrm{sech}(-8t-4),\,q(3,t)=2\mathrm{sech}(-8t+8),\,t\in[-0.5,0.5].
\end{split}
\end{align}

Utilizing the initial and boundary conditions \eqref{Ne-v2-mKdV-3}, we select $N_{\mathrm{ib}}=400$ initial-boundary points and $N_{\mathrm{c}}=10000$ collocation points to generate the training data set for LPNN-v2. We successfully trained and obtained the data-driven single-soliton solution of the mKdV equation with high accuracy using LPNN-v2, here the relative $L^2$ norm error of the solution $u(x,t)$ is 7.578669$\rm e$-04, the learned spectral parameter is $\lambda=0.056701$, and the training time and loss function iteration times of the network are 2396.8047s and 4150 times, respectively.

\begin{figure*}[!tb]
\centering
\subfigure[]{\includegraphics[height=1.2in,width=2.4in]{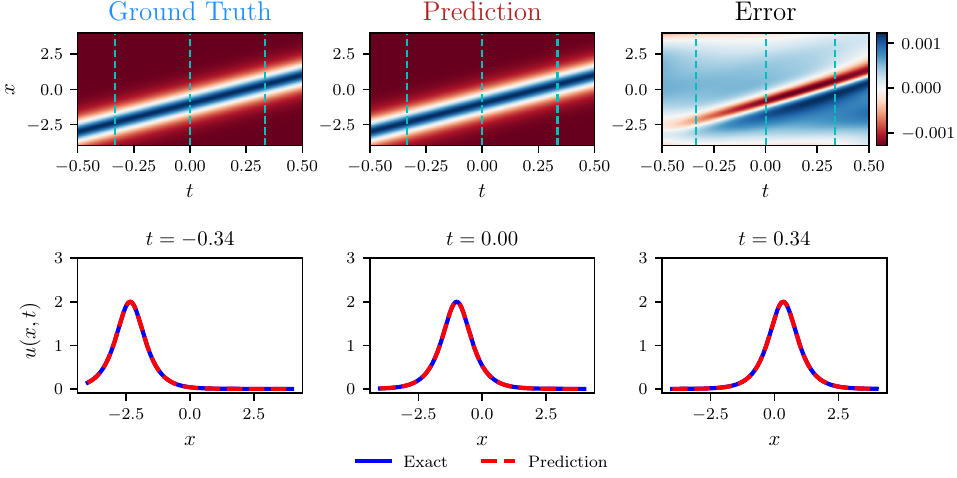}}\hspace{0.5cm}
\subfigure[]{\includegraphics[height=1.2in,width=3.0in]{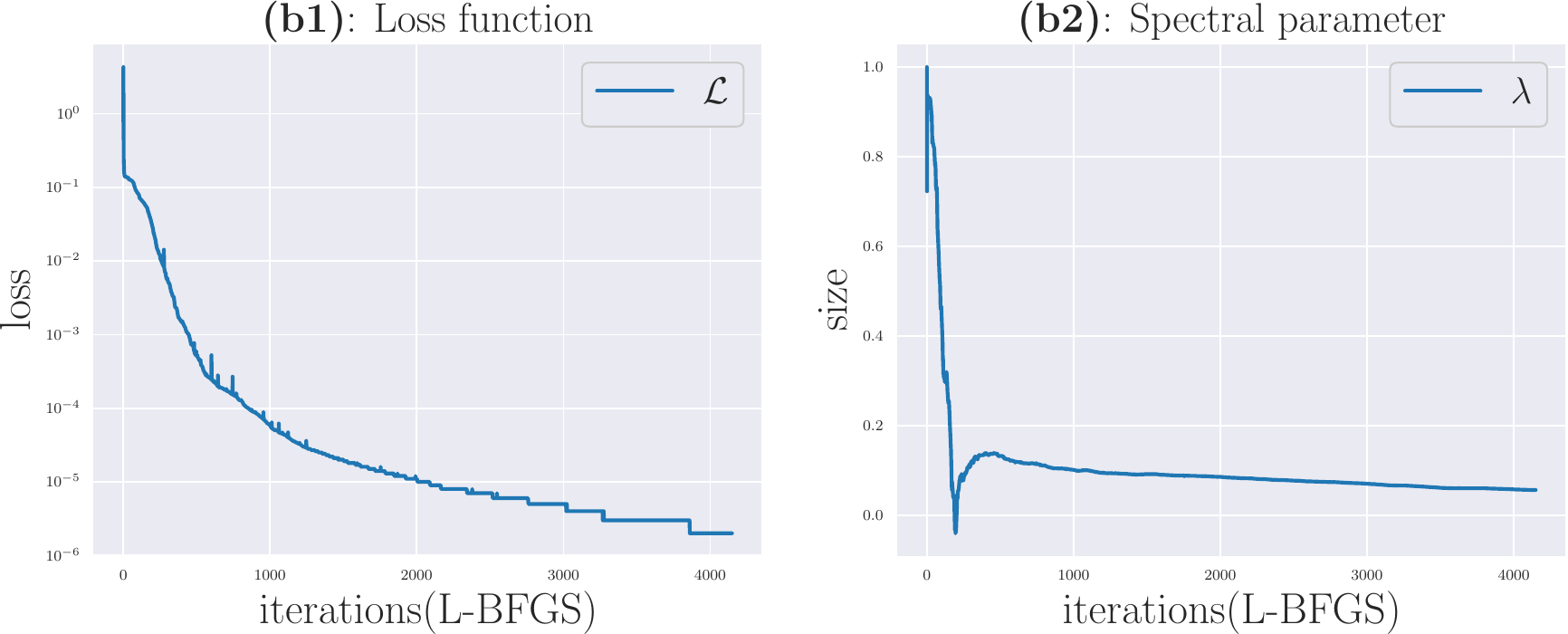}}\\
\subfigure[]{\includegraphics[height=1.0in,width=5.8in]{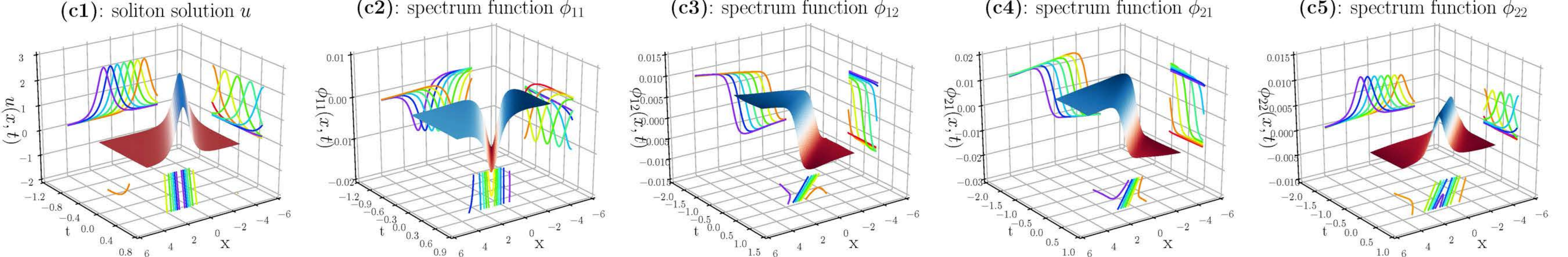}}
\caption{(Color online) The training results of single-soliton solution $u(x,t)$ and spectral functions $\phi_{i,j}(x,t)$ for mKdV equation arising from the LPNN-v2. (a) The ground truth, prediction and error dynamics density plots, as well as sectional drawings which contain the true and prediction soliton solution at three distinct moments $t=-0.34, 0, 0.34$; (b) Evolution graphs of the loss function [panel b1] and spectral parameter $\lambda$ [panels b2] in LPNN-v2; (c) The three-dimensional plots with contour map for the data-driven soliton solution [panel c1] and spectral functions [panels c2-c5] corresponding to spectral parameter $\lambda=0.056701$. \label{figLPNN-v2-mKdV}}
\end{figure*}

We present the vivid numerical results of LPNN-v2 for solving the mKdV equation in Fig. \ref{figLPNN-v2-mKdV}. In Fig. \ref{figLPNN-v2-mKdV}(a), we display detailed density plots for the ground truth, prediction and error dynamics, as well as showcase the sectional drawings at three distinct moments corresponding to the blue dashed line in density plots. The evolution graphs of the loss function [panel b1] and spectral parameter $\lambda$ [panels b2] are revealed in Fig. \ref{figLPNN-v2-mKdV}(b). Fig. \ref{figLPNN-v2-mKdV}(c1) displays the three-dimensional plot with contour map of the data-driven soliton solution $u(x,t)$, while Figs. \ref{figLPNN-v2-mKdV}(c2-c5) display the spectral functions corresponding to spectral parameter $\lambda=0.056701$. Moreover, we provide a detailed performance comparison of using three different algorithms to solve the mKdV equation in Tab. \ref{TabLPNN-v2-mKdV}. From Tab. \ref{TabLPNN-v2-mKdV}, owing to the complexity of the Lax pairs of mKdV equation, LPNN-v1 performs poorly in terms of training efficiency and accuracy. If readers prefer to use LPNN-v1, the accuracy and efficiency of LPNN-v1 can be improved by adjusting the appropriate training spatiotemporal region, resetting the hyper-parameters of NN and suitable initialization of spectral parameter. Although LPNN-v2 consumes more training time than PINN, it improves training accuracy by an order of magnitude.

\begin{table}[htbp]
  \caption{Performance comparison between LPNN-v2, LPNN-v1 and conventional PINN for solving mKdV equation}
  \label{TabLPNN-v2-mKdV}
  \centering
  \scalebox{0.8}{
  \begin{tabular}{l|c|c|c|c|c|c}
  \toprule
  \small{\textbf{Types}} & $\bm{\mathrm{x}}\times t$ & $\mathcal{D}_{\mathrm{ib}},\mathcal{D}_{\mathrm{c}}$ & optimizer & $\lambda$ & $L^2$ norm error & training time \\
  \hline
  PINN   & [-4,4]$\times$[-0.5,0.5] & 400,10000 & L-BFGS & $\backslash$ & 8.767018$\rm e$-03 & 447.3520s \\
  \hline
  LPNN-v1 & [-4,4]$\times$[-0.5,0.5] & 400,10000 & L-BFGS & 0.455595 & 7.656110$\rm e$-01 & 667.0884s \\
  \hline
  LPNN-v2 & [-4,4]$\times$[-0.5,0.5] & 400,10000 & L-BFGS & 0.056701 & 7.578669$\rm e$-04 & 2396.8047s \\
  \bottomrule
  \end{tabular}}
\end{table}

$\bullet$ \textbf{Case 3: Nonlinear Schr\"odinger equation}

The integrable NLS equation is closely related to many nonlinear problems in fields such as nonlinear optics, plasma and Bose-Einstein condensates. Unlike the integrable models studied previously, the potential function of the NLS equation is a complex valued function, and it is also one of the most classic integrable models in integrable systems, playing a crucial role in the integrable systems theory. The NLS equations have attracted much attention after Zakharov and Shabat in 1972 constructed the matrix Lax pairs and studied the inverse scattering transform and exact solutions for the NLS equation \cite{Zakharov-SPJ-1972}. The NLS equation with two auxiliary linear equations [Lax pairs] can be written as
\begin{align}\label{Ne-v2-NLS-1}
\mathrm{i}q_t+\frac12q_{xx}+|q|^2q=0,
\end{align}
\begin{align}\label{Ne-v2-NLS-2}
f_{\rm{Lp}}:\,\bigg\{\begin{aligned}
\Phi_{x}=M\Phi\\
\Phi_t=N\Phi
\end{aligned},\,
M=\begin{bmatrix} \mathrm{i}\lambda & \mathrm{i}q^*  \\ \mathrm{i}q & -\mathrm{i}\lambda \end{bmatrix},\,N=\begin{bmatrix} \mathrm{i}\lambda^2-\frac12\mathrm{i}qq^* & \mathrm{i}\lambda q^*+\frac12q^*_x  \\ \mathrm{i}\lambda q-\frac12q_x & -\mathrm{i}\lambda^2+\frac12\mathrm{i}qq^* \end{bmatrix},
\end{align}
where the superscript ``*'' represents complex conjugation, $|q|$ indicate modulus of complex-value solution $q(x,t)\in\mathbb{C}^{1\times1}$, then we set $q(x,t)=u(x,t)+\mathrm{i}v(x,t),\,u(x,t)/v(x,t)\in\mathbb{R}^{1\times1}$ for network training. Similarly, spectral function $\Phi(x,t)\in\mathbb{C}^{2\times 1}$, thus we let $\Phi(x,t)=(\phi_1(x,t),\phi_2(x,t))^{\rm T}$ and $\phi_1(x,t)=\phi_{11}(x,t)+\mathrm{i}\phi_{12}(x,t),\phi_2(x,t)=\phi_{21}(x,t)+\mathrm{i}\phi_{22}(x,t)$, in which $\phi_{ij}(x,t)\in\mathbb{R}^{1\times 1}[i,j=1,2]$. Furthermore, spectral parameter $\lambda\in\mathbb{C}$, then we take $\lambda=\lambda_1+\lambda_2\mathrm{i}$, here we have $\lambda_1\in\mathbb{R}$ and $\lambda_2\in\mathbb{R}$. The NLS equation \eqref{Ne-v2-NLS-1} can be derived via the Lax pairs \eqref{Ne-v2-NLS-2} and zero curvature equation \eqref{I5}. Next we initialize $\phi_{ij}=0$ and make it satisfy the free initial-boundary conditions. Then according to Ref. \cite{PuJC-CPB-2021,Akhmediev-PRE-2009}, we provide the following initial-boundary value conditions of NLS equation in spatiotemporal region $[-3,3]\times[-1.5,1.5]$
\begin{align}\label{Ne-v2-NLS-3}
\begin{split}
&q(x,t=-1.5)=\bigg[1-\frac{4(1-3\mathrm{i})}{4x^2+10}\bigg]\mathrm{e}^{-1.5\mathrm{i}},\,x\in[-3,3],\\
&q(-3,t)=\bigg[1-\frac{4(1+2\mathrm{i}t)}{4t^2+37}\bigg]\mathrm{e}^{\mathrm{i}t},\,q(3,t)=\bigg[1-\frac{4(1+2\mathrm{i}t)}{4t^2+37}\bigg]\mathrm{e}^{\mathrm{i}t},\,t\in[-1.5,1.5].
\end{split}
\end{align}

Likewise, according to the initial and boundary conditions \eqref{Ne-v2-NLS-3}, we choose $N_{\mathrm{ib}}=400$ initial and boundary points, as well as $N_{\mathrm{c}}=10000$ collocation points in LPNN-v2. The complex-value spectral parameter $\lambda$ is initialized to $-0.5+0.5\mathrm{i}$, then the relative $L^2$ norm error reaches 1.264692$\rm e$-03 for data-driven rogue wave solution $q(x,t)$ by applying 13733 L-BFGS optimization in 6446.000261 seconds. Meanwhile, the spectral parameter of numerical discovery is $\lambda=0.38823+0.018209\mathrm{i}$, and the corresponding vector spectral function $\Phi$ in spectral problem \eqref{Ne-v2-NLS-2} has learned with high accuracy from LPNN-v2. Rogue wave is a very rare and short-lived isolated large amplitude wave, characterized by "coming without a shadow, going without a trace", and its wave height is generally more than twice that of the surrounding highest wave \cite{Kharif-EJMB-2003}. The NLS equation \eqref{Ne-v2-NLS-1} is also the most basic mathematical model for describing rogue wave phenomena. Hitherto, rogue waves have emerged in numerous research fields, such as optics and fluid mechanics \cite{Solli-N-2007,Chabchoub-PRL-2011}.

\begin{figure*}[!tb]
\centering
\subfigure[]{\includegraphics[height=1.0in,width=2.0in]{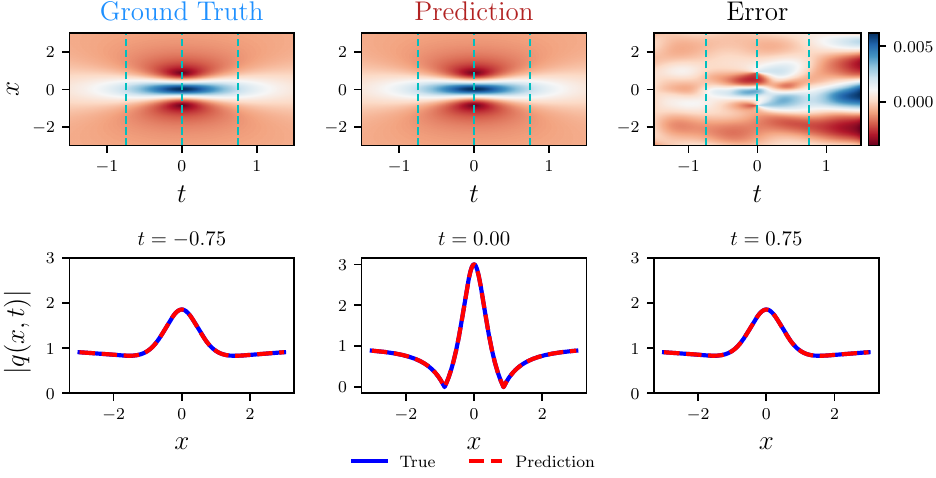}}\hspace{0.5cm}
\subfigure[]{\includegraphics[height=1.0in,width=3.6in]{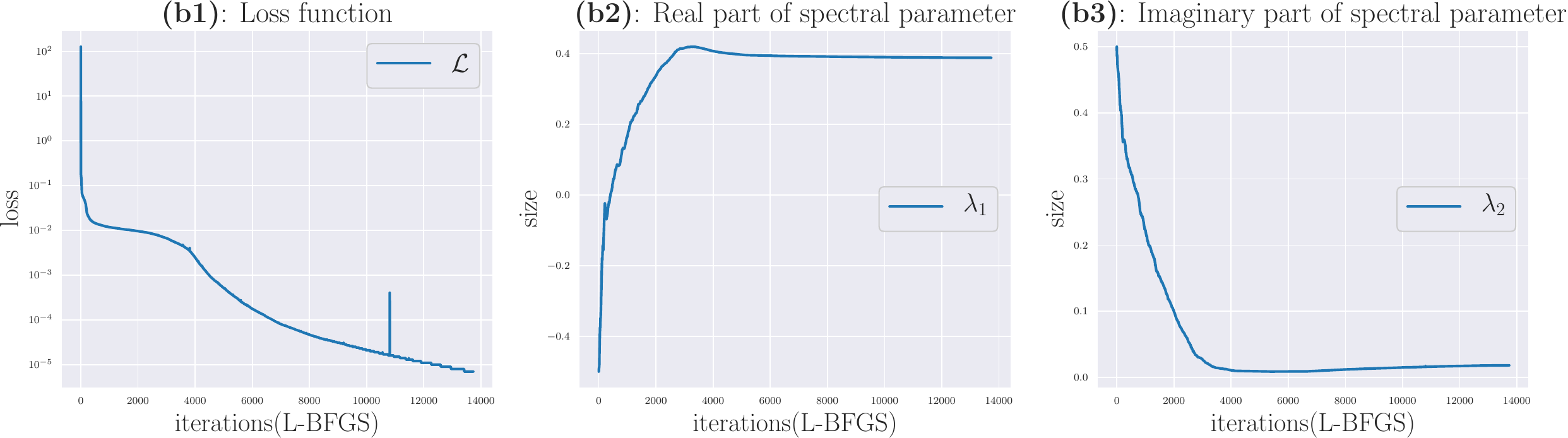}}\\
\subfigure[]{\includegraphics[height=1.4in,width=5.8in]{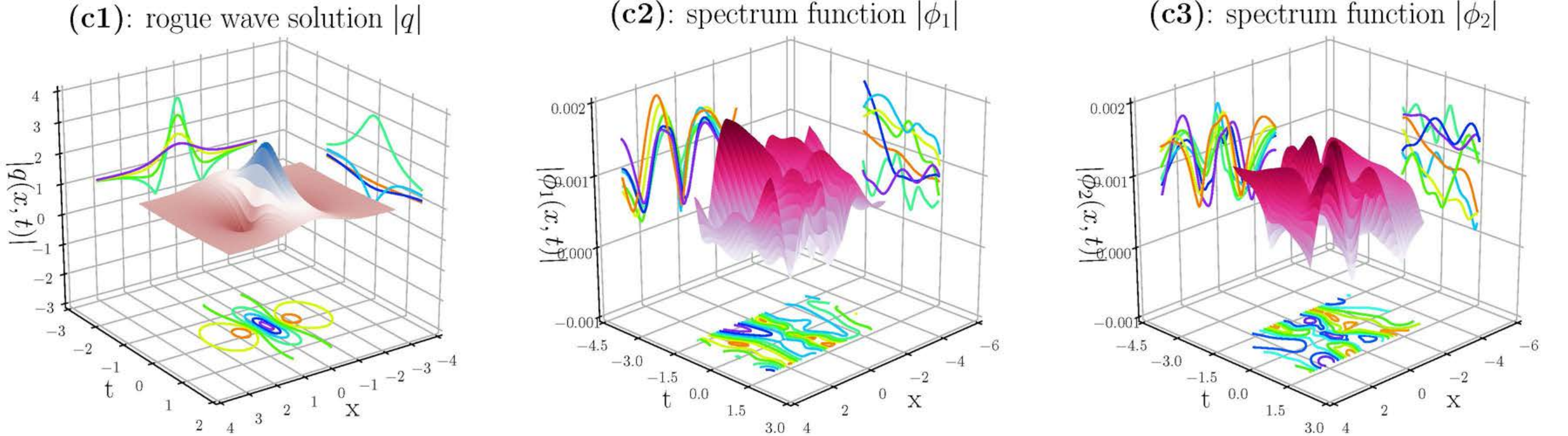}}
\caption{(Color online) The training results of rogue wave solution $|q(x,t)|$ and spectral functions $|\phi_1(x,t)|,|\phi_2(x,t)|$ for NLS equation arising from the LPNN-v2. (a) The ground truth, prediction and error dynamics density plots, as well as sectional drawings which contain the true and prediction rogue wave solution at three distinct moments $t=-0.75, 0, 0.75$; (b) Evolution graphs of the loss function [panel b1] and spectral parameter $\lambda=\lambda_1+\mathrm{i}\lambda_2$ [panels b2, b3] in LPNN-v2; (c) The three-dimensional plots with contour map for the data-driven rogue wave solution [panel c1] and spectral functions [panels c2, c3] corresponding to spectral parameter $\lambda=-0.046563+0.257338\mathrm{i}$. \label{figLPNN-v2-NLS}}
\end{figure*}

Fig. \ref{figLPNN-v2-NLS} displays the training results of the data-driven rogue wave solution and corresponding spectral problem for the NLS equation stemming from the LPNN-v2. Similarly, the abundant density plots and sectional drawings are revealed in Fig. \ref{figLPNN-v2-NLS}(a), then Fig. \ref{figLPNN-v2-NLS}(b) reveals the loss function curve [panel (b1)] and the numerical evolution of the real and imaginary parts of the spectral parameters $\lambda$ [panels (b2) and (b3)]. The three-dimensional plots and its contour map on three planes for the data-driven rogue wave solution and spectral functions are exhibited in Fig. \ref{figLPNN-v2-NLS}(c). We provide the performance comparison of utilizing three different methods to solve the NLS equation in Tab. \ref{TabLPNN-v2-NLS}. From Tab. \ref{TabLPNN-v2-NLS}, we can observe that LPNN-v2 has the highest training accuracy after consuming more training time, while LPNN-v1 has the shortest training time but lower training accuracy. Compared with traditional PINN, LPNN-v2 has improved training accuracy by more than twice. Certainly, we can further improve the training accuracy of LPNN-v1 and LPNN-v2 by readjusting the training spatiotemporal region, resetting hyper-parameters and spectral parameters.

\begin{table}[htbp]
  \caption{Performance comparison between LPNN-v2, LPNN-v1 and conventional PINN for solving NLS equation}
  \label{TabLPNN-v2-NLS}
  \centering
  \scalebox{0.8}{
  \begin{tabular}{l|c|c|c|c|c|c}
  \toprule
  \small{\textbf{Types}} & $\bm{\mathrm{x}}\times t$ & $\mathcal{D}_{\mathrm{ib}},\mathcal{D}_{\mathrm{c}}$ & optimizer & $\lambda$ & $L^2$ norm error & training time \\
  \hline
  PINN   & [-3,3]$\times$[-1.5,1.5] & 100,10000 & L-BFGS & $\backslash$ & 3.375881$\rm e$-03 & 2911.307015s \\
  \hline
  LPNN-v1 & [-3,3]$\times$[-1.5,1.5] & 100,10000 & L-BFGS & -0.374518+0.393163$\mathrm{i}$ & 4.112298$\rm e$-01 & 979.621199s \\
  \hline
  LPNN-v2 & [-3,3]$\times$[-1.5,1.5] & 100,10000 & L-BFGS & 0.38823+0.018209$\mathrm{i}$ & 1.264692$\rm e$-03 & 6446.000261s \\
  \bottomrule
  \end{tabular}}
\end{table}

$\bullet$ \textbf{Case 4: Short pulse equation}

The short pulse (SP) equation
\begin{align}\label{Ne-v2-SP-1}
u_{xt}=u+\frac16(u^3)_{xx}
\end{align}
is proposed by Sch\"afer and Wayne \cite{Schafer-PD-2004} as an alternative (to the NLS eqaution) model for approximating the evolution of ultrashort intense infrared pulses in silica optical, in which $u$ represents the dimensionless electric field, parameters $t$ and $x$ standing for time-like and space-like independent variables, respectively. Subsequent numerical analyses have demonstrated that the SP equation as an approximation of the Maxwell's equation has superior applicability, particularly when characterizing the propagation of extremely short-duration light pulses \cite{Chung-N-2005}. This equation was initially introduced as one of Rabelo's equations in the realm of differential geometry \cite{Rabelo-SAM-1989}. Furthermore, aforementioned KdV equation, mKdV equation, SG equation and NLS equation all belong to the AKNS hierarchy \cite{Ablowitz-PRL-1973}, while the SP equation is the most classical integrable model in WKI equations and possesses a Lax pairs of the WKI type \cite{Wadati-JPSJ-1979}, and it can be transformed into the integrable SG equation \cite{Sakovich-JPSJ-2005}. One can obtain the Lax pairs of SP equation, as shown in following
\begin{align}\label{Ne-v2-SP-2}
f_{\rm{Lp}}:\,\bigg\{\begin{aligned}
\Phi_{x}=M\Phi\\
\Phi_t=N\Phi
\end{aligned},\,
M=\begin{bmatrix} \lambda & \lambda u_x  \\ \lambda u_x & -\lambda \end{bmatrix},\,N=\begin{bmatrix} \frac12\lambda u^2+\frac{1}{4\lambda} & \frac12\lambda u^2u_x-\frac12u  \\ \frac12\lambda u^2u_x+\frac12u & -\frac12\lambda u^2-\frac{1}{4\lambda} \end{bmatrix}.
\end{align}
The SP equation \eqref{Ne-v2-SP-1} can be derived by utilizing the zero curvature equation \eqref{I5} along with Lax pairs \eqref{Ne-v2-SP-2}. Similar to case 3, we set spectral function $\Phi(x,t)\in\mathbb{C}^{2\times 1}$ and spectral parameter $\lambda\in\mathbb{C}$, thus we take $\lambda=\lambda_1+\lambda_2\mathrm{i}$, $\Phi(x,t)=(\phi_1(x,t),\phi_2(x,t))^{\rm T}$ and $\phi_1(x,t)=\phi_{11}(x,t)+\mathrm{i}\phi_{12}(x,t),\phi_2(x,t)=\phi_{21}(x,t)+\mathrm{i}\phi_{22}(x,t)$, where $\lambda_1/\lambda_2\in\mathbb{R}$ and $\phi_{ij}(x,t)\in\mathbb{R}^{1\times 1}[i,j=1,2]$. Next we initialize $\phi_{ij}=0$ and make it satisfy the free initial-boundary conditions, then we refer to Ref. \cite{Sakovich-JPAMG-2006} and consider the initial and boundary value conditions of SP equation in spatiotemporal region $[-6,6]\times[-3,3]$, as shown in following
\begin{align}\label{Ne-v2-SP-3}
\begin{split}
&q(x,t=-3)=\frac{0.04\cos(x+3)}{\cosh(0.01x-0.03))},\,x\in[-6,6],\\
&q(-6,t)=\frac{0.04\cos(-6-t)}{\cosh(-0.06+0.01t)},\,q(6,t)=\frac{0.04\cos(6-t)}{\cosh(0.06+0.01t)},\,t\in[-3,3].
\end{split}
\end{align}

From aforementioned initial and boundary value conditions \eqref{Ne-v2-SP-3}, we select $N_{\rm ib}=400$ and $N_{\rm c}=10000$ for training of LPNN-v2, while the spectral parameter was initialized as $\lambda=0.5+0.5\mathrm{i}$. After 5913 L-BFGS optimization in LPNN-v2, the relative $L^2$ norm error reaches 1.528543$\rm e$-02 in 2400.7532 seconds, and the optimal spectral parameter learned is $\lambda=0.002629+0.047998\mathrm{i}$. As shown in Ref. \cite{Sakovich-JPAMG-2006}, the data-driven solution learned is been called the pulse solution, it represents a single-valued nonsingular pulse or a wave packet, and the shape of the pulse is similar to that of the NLS soliton: the $\mathrm{sech}$-shaped envelope modulates the $\cos$-shaped wave. Moreover, owing to the initial and boundary training points in our training data come from approximate pulse solution of the SP equation \eqref{Ne-v2-SP-1}, we can find that the $L^2$ error trained by LPNN is relatively large. Based on the high-precision characteristics of LPNN-v2, we have reason to believe that the pulse solution learned in this part is more realistic.

\begin{figure*}[!tb]
\centering
\subfigure[]{\includegraphics[height=1.0in,width=2.0in]{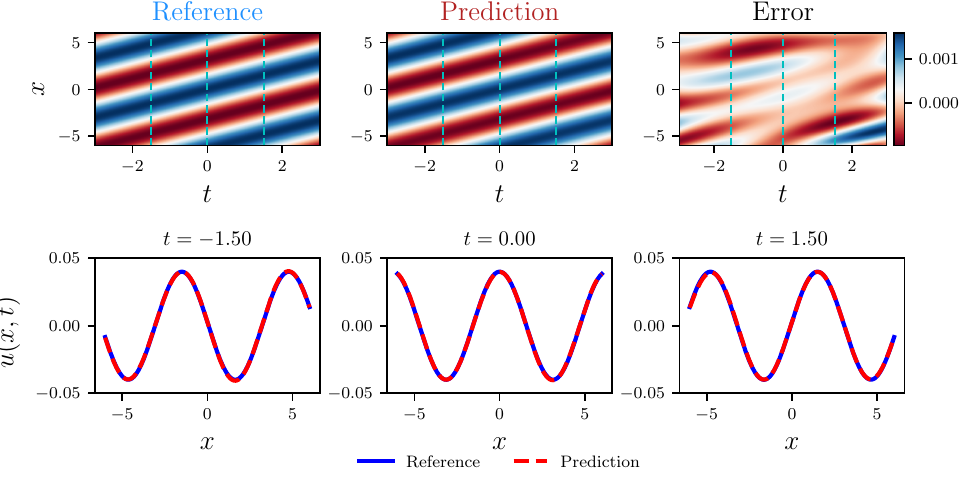}}\hspace{0.5cm}
\subfigure[]{\includegraphics[height=1.0in,width=3.6in]{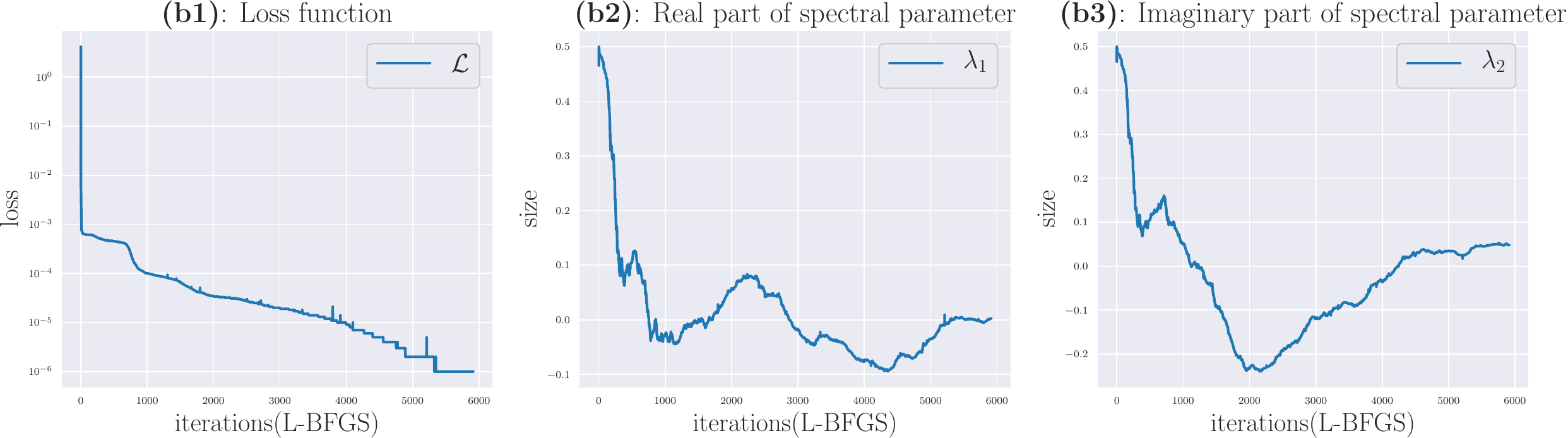}}\\
\subfigure[]{\includegraphics[height=1.4in,width=5.8in]{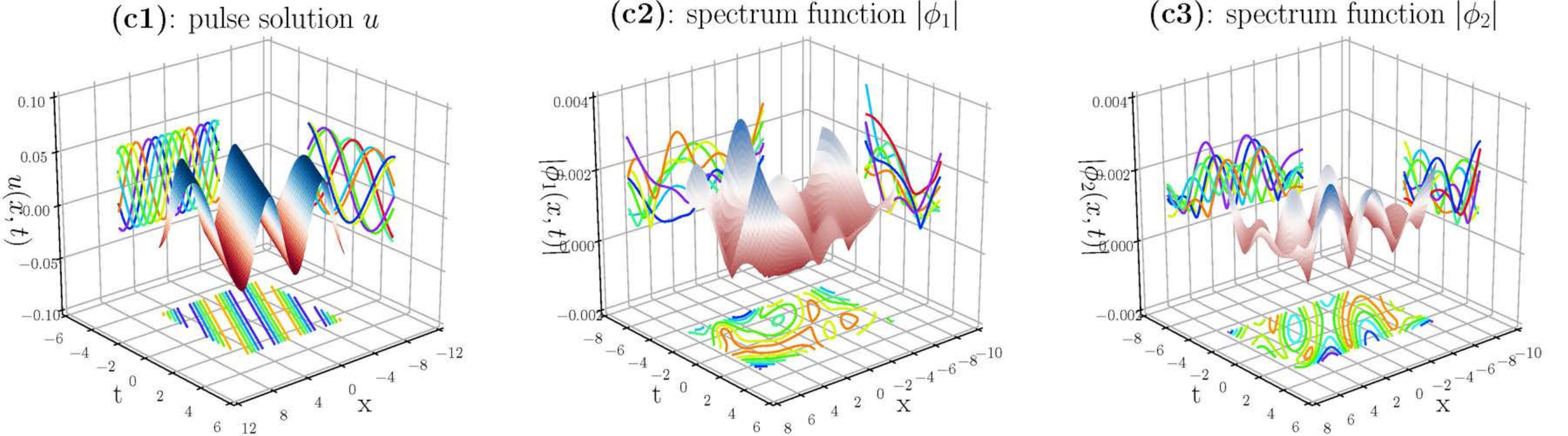}}
\caption{(Color online) The training results of pulse solution $u(x,t)$ and spectral functions $\phi_1(x,t),\phi_2(x,t)$ for SP equation arising from the LPNN-v2. (a) The ground truth, prediction and error dynamics density plots, as well as sectional drawings which contain the reference and prediction pulse solution at three distinct moments $t=-1.5, 0, 1.5$; (b) Evolution graph of the loss function [panel b1] and spectral parameter [panels b2, b3] in LPNN-v2; (c) The three-dimensional plots with contour map for the data-driven pulse solution [panel c1] and spectral functions [panels c2, c3] corresponding to spectral parameter $\lambda=0.002629+0.047998\mathrm{i}$. \label{figLPNN-v2-SP}}
\end{figure*}

The training results of the data-driven pulse solution stemming from the LPNN-v2 are indicated in Fig. \ref{figLPNN-v2-SP}, in which Fig. \ref{figLPNN-v2-SP}(a) displays the abundant density plots and sectional drawings. Fig. \ref{figLPNN-v2-SP}(b) shows the loss function curve in panel (b1), and the numerical evolutions for the real and imaginary parts of the spectral parameter $\lambda$ with network training are shown in (b2) and (b3), respectively. While Fig. \ref{figLPNN-v2-SP}(c) showcases the three-dimensional plot and its contour map on three planes for pulse solution $u(x,t)$ and spectral functions $\phi_1,\phi_2$. Tab. \ref{TabLPNN-v2-SP} exhibits a detailed comparison of training results by applying three algorithms to solve the SP equation. Similar to the numerical results of solving the NLS equation in case 3, LPNN-v2 has improved training accuracy by more than twice compared to the traditional PINN algorithm. Due to the fact that the Lax pairs of the SP equation are more complicated compared to the SP equation itself, so LPNN-v1 has poor training accuracy in solving the SP equation. Therefore, as solving this type of integrable systems with intricate Lax pairs, we usually recommend using LPNN-v2 to obtain higher training accuracy.

\begin{table}[htbp]
  \caption{Performance comparison between LPNN-v2, LPNN-v1 and conventional PINN for solving SP equation}
  \label{TabLPNN-v2-SP}
  \centering
  \scalebox{0.8}{
  \begin{tabular}{l|c|c|c|c|c|c}
  \toprule
  \small{\textbf{Types}} & $\bm{\mathrm{x}}\times t$ & $\mathcal{D}_{\mathrm{ib}},\mathcal{D}_{\mathrm{c}}$ & optimizer & $\lambda$ & $L^2$ norm error & training time \\
  \hline
  PINN   & [-6,6]$\times$[-3,3] & 400,10000 & L-BFGS & $\backslash$ & 3.432290$\rm e$-02 & 577.9960s \\
  \hline
  LPNN-v1 & [-6,6]$\times$[-3,3] & 400,10000 & L-BFGS & 0.037306+0.040709$\mathrm{i}$ & 1.664159$\rm e$+00 & 532.3504s \\
  \hline
  LPNN-v2 & [-6,6]$\times$[-3,3] & 400,10000 & L-BFGS & 0.002629+0.047998$\mathrm{i}$ & 1.528543$\rm e$-02 & 2400.7532s \\
  \bottomrule
  \end{tabular}}
\end{table}

In this subsection, we further introduce compatibility condition/zero curvature equation constraints into Lax pairs informed based on LPNN-v1, then propose LPNN-v2 that can achieve higher accuracy. We investigated the data-driven localized wave solutions and spectral problems of several important integrable systems by means of LPNN-v2, and found that LPNN-v2 achieved higher training accuracy than standard PINN and LPNN-v1 by sacrificing training time. Specifically, we employed the LPNN-v2 to solve data-driven localized wave solutions and their Lax pairs spectral problems for the SG equation, mKdV equation, NLS equation and SP equation, where localized wave solutions include kink solution, soliton solution, rogue wave solution and pulse solution. The numerical results indicate that the LPNN-v2 has a wider range of applications than LPNN-v1, and can usually be utilized to study all integrable systems with Lax pairs with high accuracy.

\section{Conclusions}
In this article, we introduce Lax pairs, which is the most important feature of integrable systems, into deep NN algorithms and propose the LPNNs suitable for integrable systems. LPNN-v1 relies entirely on the Lax pairs of integrable systems and introduces Lax pairs informed into the loss function, which can efficiently study the localized wave solutions and spectral problems of certain integrable systems. Based on the LPNN-v1, the Lax pairs constraint and loss function of LPNN-v2 also depend on the compatibility condition equation or zero curvature equation of the integrable system, which can accurately solve all integrable systems with high-accuracy and obtain the corresponding spectral parameter and spectral function in the Lax pairs spectral problem.

This article designs several deep learning numerical experiments for important integrable systems with Lax pairs. Specifically, the KdV equation, CH equation, KP equation and high-dimensional KdV equation were efficiently solved by utilizing LPNN-v1, and the data-driven soliton solution, non-smooth peakon solution, line-soliton solution, lump solution, spectral parameter and corresponding spectral function were learned and obtained, here the training time of LPNN-v1 was more than two to five times faster than that of the standard PINN. The data-driven localized wave solutions and their corresponding spectral problems of the SG equation, mKdV equation, NLS equation and SP equation were studied with high precision by means of LPNN-v2, in which the data-driven localized wave solutions include kink solution, soliton solution, rogue wave solution and pulse solution. LPNN-v2 achieved higher accuracy than standard PINN and LPNN-v1, and it is suitable for all integrable systems with Lax pairs. Generally, setting different spectral parameter initialization in LPNNs often leads to different spectral function, training time and training accuracy, which often need to be adjusted based on results and experience in practical numerical experiments.

By means of replacing or assisting the PDE in Lax pairs informed network, we introduce the most important feature of integrable systems, namely Lax pairs, into deep NNs in the first time. Then we propose novel LPNNs suitable for the integrable systems with Lax pairs, in which LPNN-v1 can attain efficient training performance, while LPNN-v2 can achieve high training accuracy. The study of numerical Lax pairs spectral problems in integrable systems using deep learning methods starting from Lax pairs has not yet been reported, so the research results of this work will be highly distinctive and unique. The research work of this article will further promote the development of the framework of integrable deep learning methods, providing new ideas and approaches for the numerical spectral problems of integrable systems, the discovery of new localized wave solutions and even new Lax pairs.

\section*{Declaration of competing interest}
The authors declare that they have no known competing financial interests or personal relationships that could have appeared to influence the work reported in this paper.

\section*{Data availability}
Data will be made available on request.

%\section*{Acknowledgements}
%\hspace{0.3cm}
%The authors gratefully acknowledge the support of the National Natural Science Foundation of China (No. 12175069 and No. 12235007), Science and Technology Commission of Shanghai Municipality (No. 21JC1402500 and No. 22DZ2229014), and Natural Science Foundation of Shanghai (No. 23ZR1418100).

%\section*{Acknowledgements}
%\hspace{0.3cm}
%This work was supported by the Sino-German Science Center (Grant No. GZ 1465) for S. Jiang, by the CAEP foundation (No. CX20200026) and National Key Project (GJXM92579) for W.J. Sun, by the National Natural Science Foundation of China (No. 12175069 and No. 12235007) for Y. Chen.

\end{document}